\documentclass[aps,pra,twocolumn,showpacs]{revtex4-1}
\usepackage {amsmath}
\usepackage {amssymb}
\usepackage[dvips]{graphicx}
\usepackage[utf8]{inputenc}
\usepackage[T2A]{fontenc}
\usepackage[english]{babel}
\usepackage{indentfirst}

\usepackage[usenames]{color}
\usepackage{ulem}
\usepackage[colorlinks=true]{hyperref}

\newcommand{\rmi}{\mathrm{i}}
\newcommand{\rme}{\mathrm{e}}

\renewcommand{\vec}[1]{\boldsymbol{#1}}
\newcommand{\Imag}{\mathop{\mathcal{I}m}}
\newcommand{\ie}{\textit{i.e.},~}

\begin{document}
 
\title{Self-compression of soliton-like laser pulses in the process of self-focusing}

\author{A.~A.~Balakin}
\author{A.~G.~Litvak}
\author{V.~A.~Mironov}
\author{S.~A.~Skobelev}
\email{sksa@ufp.appl.sci-nnov.ru}

\affiliation{Institute of Applied Physics of the Russian Academy of Sciences, Nizhny Novgorod, Russia}

\date{\today}

\begin{abstract}
We study the possibility of efficient self-compression of femtosecond laser pulses in nonlinear media with anomalous dispersion of group velocity during the self-focusing of wave packets with a power several times greater than the critical self-focusing power. The results of qualitative analysis of the evolution of three-dimensional wave packets with the quasi-soliton field distribution are confirmed by the computer simulation. The simulation proves that the considered regime of compression of high-power laser pulses with initial durations of about ten optical cycles is stable relative to filamentation instability due to the influence of the nonlinear dispersion. We demonstrate the possibility of self-compression of laser pulses at a multi-millijoule energy level and up to one optical cycle with an energy efficiency of more then 50\%.
\end{abstract}

\pacs{42.65.-k, 42.50.-p, 42.65.Jx} 
\maketitle
\section{Introduction}

The generation of high-energy few-cycle optical pulses is a challenge in contemporary laser physics that has important implications for high-field science and as well for many other extreme light applications \cite{krausz}. Conventional laser systems, based on broadband active media, e.g., such as Ti:sapphire crystals and/or parametric amplification technology, are able to provide pulses of ultrashort durations with high enough energies. However, to get high-energy pulses of few-cycle or even single cycle duration require additional spectral broadening with subsequent compression technique. For example, usual way is to broaden spectrum of the pulse in a high pressure gas and then to compress it by using gratings or chirp mirrors. Alternative but more refined way is to use self-compression mode employing Kerr nonlinearity \cite{kivshar,liu}, for high-power pulses ionization~\cite{wagner,kim,ionization} and relativistic~\cite{wake} nonlinearities. It should be noted two important features on which this way relies. First, anomalous dispersion is required to realize self-compression mode. Second,  waveguiding of pulse propagation is used for long interaction distances, which imposes some restrictions on the pulse energy. It is worth to emphasize that at least for powers about the critical power for self-focusing $P_\text{cr}$ when single filamentation occurs, self-guided pulses can be stable and their self-compressing favorably used for obtaining pulses of extremely short durations \cite{chin,faccio,mysyrowicz}. Of course, self-compression of optical pulses in anomalous group velocity dispersion media with the help of transversely cummulating laser energy, \ie self-focusing, looks very attractive idea for getting high-energy few-cycle pulses. Nevertheless, for high energy pulses with peak power much exceeding the threshold of self-focusing, the filamentation instability is the main obstacle that completely prevents using free space propagation in nonlinear bulk media.

It should be noted that most of the media in the visible frequency range are characterized by normal dispersion of group velocities. However, in recent years, considerable success has been achieved in the field of generation of high-energy laser pulses in the mean IR range. In this frequency range, most of the media are characterized by anomalous group velocity dispersion. The interest in ultrashort high-energy laser pulses in the $2$--$8$-$\mu$m range is directly related to their specific features connected with generation of higher harmonics~\cite{Kapteyn,Corkum}, filamentation~\cite{Silva,Shim}, and particle acceleration. 

Numerical simulation and experiments demonstrated that filamentation of optical radiation in the mean IR range follows a different scenario~\cite{chin,Naudeau,Bradler}. A peculiar feature of filamentation dynamics is possible formation of a set of soliton structures in the optical field, which have rather short durations (two or three field periods)~\cite{mysyrowicz}. This regime of laser pulse shortening is limited by the powers about of the critical self-focusing power. This take place due to development of the filamentation instability, which leads to fast decomposition of the beam to many filaments, and the final loss of coherence makes the sources of such pulses inapplicable in practice. 

In this work, basing on the results described in~\cite{Skobelev16}, we present detailed analytical and numerical studies, which demonstrate that nonlinear dispersion of the medium (dependence of the group velocity of the wave packet on its intensity) leads to stabilization of filamentation instability for ultrashort pulses. It is shown that at anomalous medium dispersion, the filamentation-free regime of laser pulse self-focusing allows one to realize self-compression up to one optical cycle for soliton-like pulses having powers much larger than critical one for the self-focusing.

The structure of this work is as follows. In Section~\ref{sec:2}, the basic equation is formulated for describing self-action of ultrashort laser pulses comprising several optical cycles in a medium with a Kerr-type inertia-free nonlinearity. Sections~\ref{sec:3} and \ref{sec:4} present a qualitative study of the method for adiabatic compression of the laser pulse duration. In Section~\ref{sec:5}, stability of the radiation self-focusing is studied in relation to the filamentation instability. In the last part, Section\ref{sec:6}, the results of numerical simulation are presented, and the initial parameters of the laser pulse for optimal self-compression of the wave packet are determined. 

\section{Basic equations}\label{sec:2}
To describe adequately the spatio-temporal evolution of ultrashort, circularly polarized laser pulses ($\vec{E}=\mathcal{E}(\vec{x}_0+\rmi \vec{y}_0)$, where $\mathcal{E}_x$ and $\mathcal{E}_y$ are the corresponding components of the electric-field intensity) in a medium with cubic nonlinearity at the self-focusing process. Let us turn now directly to the wave equation
\begin{equation}\label{eq:1}
\dfrac{\partial^2\mathcal{E}}{\partial z^2}+\Delta_\perp \mathcal{E}-\dfrac{1}{c^2}\int\limits_{-\infty}^{t}\varepsilon(t-t')\mathcal{E}(t')dt'=\dfrac{4\pi}{c^2}\dfrac{\partial^2\mathcal{P}_\text{nl}}{\partial t^2} .
\end{equation}
Here, $\mathcal{P}_\text{nl}$ is the nonlinear medium response, and $\varepsilon$ is the linear dielectric permittivity, which satisfies the fundamental Kramers---Kronig relation
\begin{equation}\label{eq:2}
\varepsilon_r(\omega)=1+\dfrac{2}{\pi}\int\limits_{0}^{\infty}\dfrac{x\varepsilon_i(x)dx}{x^2-\omega^2} ,
\end{equation}
where
$\varepsilon_r$ and $\varepsilon_i$ are the real and imaginary parts of the dielectric permittivity $\varepsilon$. Let us apply Eq.~\eqref{eq:2} to weakly absorbing media, \ie assume that one can neglect the imaginary part of the dielectric permittivity within the frequency range being of interest for us. Let's assume that the weak-absorption region spreads in a wide frequency range from $\omega_1$ to $\omega_2$, and consider the frequencies $\omega$, such that $\omega_1\ll\omega\ll\omega_2$. 
\begin{multline}
\varepsilon_r(\omega)\simeq1-\dfrac2{\pi}\int\limits_0^{\omega_1}\dfrac{x\varepsilon_i(x)dx}{\omega^2}+\dfrac2{\pi}\int\limits_{\omega_2}^{+\infty}\dfrac{\varepsilon_i(x)dx}{x\left(1-\dfrac{\omega^2}{x^2}\right)}\simeq\\
\simeq \varepsilon_0-\dfrac{a}{\omega^2}+b\omega^2  ,
\end{multline}
where $\varepsilon_0=1+(2/\pi)\int_{\omega_2}^{\infty}(\varepsilon_i/x)dx$ is the static dielectric permittivity, 
$\omega_D^2=(2/\pi)\int_{0}^{\omega_1}x\varepsilon_idx$, and $b=(2/\pi)\int_{\omega_2}^{\infty}(\varepsilon_i/x^3)dx$. 
Then, we will limit our consideration to the case, where $\varepsilon_0\gg \omega_D^2/\omega^2\gg b\omega^2$. 
This condition is fulfilled well for the majority of media in the mean IR range. 
As a result, the dielectric permittivity will take on the following form: 
\begin{equation}
\varepsilon(\omega)\simeq\varepsilon_0-\dfrac{\omega_D^2}{\omega^2} . 
\end{equation}

In the case of a nonresonance medium with the instantaneous nonlinear response $\mathcal{P}_\text{nl}=\chi^{(3)}|\mathcal{E}|^2\mathcal{E}$ (here, $\chi^{(3)}$ is the cubic nonlinear susceptibility), wave equation~\eqref{eq:1} will acquire the following form~\cite{Anderson,Kozlov}:
\begin{equation}\label{eq:4}
\dfrac{\partial^2\mathcal{E}}{\partial z^2}+\Delta_\perp \mathcal{E}-\dfrac{\varepsilon_0}{c^2}\dfrac{\partial^2\mathcal{E}}{\partial t^2}-\dfrac{\omega_D^2}{c^2}\mathcal{E} -\dfrac{4\pi\chi^{(3)}}{c^2}\dfrac{\partial^2 |\mathcal{E}|^2\mathcal{E}}{\partial t^2}=0 .
\end{equation}
This equation describes the spatio-temporal evolution of the field intensity $\mathcal{E}$ of the laser pulse allowing for all considered important physical factors, such as medium dispersion, beam diffraction, Kerr nonlinearity, self-steepening of the wave packet, and self-focusing of the transverse distribution. 

For detailed analysis of a dynamic problem, it is convenient to use the evolution equation for the field in the simplest form of a reduced wave equation. Let's suppose that the spatio-temporal structure of the wave field varies smoothly in the process of one-way pulse propagation, \ie let's neglect the reflection effects. In the quasimonochromatic case, this approach corresponds to passing over to an envelope equation. Using the approximation of one-way propagation of the wave field along the $z$ axis, \ie supposing smallness of $|c \partial_z \mathcal{E}+\partial_t \mathcal{E}| \ll |c \partial_z \mathcal{E} - \partial_t \mathcal{E}|$, we will represent the equation in dimensionless variables in the following form~\cite{Skobelev_collaps_PRA09,Skobelev_collaps_PRA08,Anderson,Kozlov}:
\begin{equation}\label{eq:5}
\dfrac{\partial^2u}{\partial z\partial\tau}+u+\dfrac{\partial^2}{\partial\tau^2}(|u|^2u)=\Delta_\perp u .
\end{equation}
Here $u=\mathcal{E}\sqrt{4\pi\chi^{(3)}}\omega_0/\omega_D$, $z\to z2\sqrt{\varepsilon_0}\omega_0c/\omega_D^2 $, $\tau=\omega_0(t-z\sqrt{\varepsilon_0}/c)$, $\omega_0$ is the characteristic carrier frequency, and $r_\perp \to r_\perp c/\omega_D$.

In the case of the monochromatic wave packet 
$$u=\Psi(z,\tau,r_\perp )\exp(\rmi\omega_0\tau-\rmi k_z z),$$
Eq.~\eqref{eq:5} yields easily an equation, which generalizes the nonlinear Schr\"odinger equation (NSE) for the envelope~\cite{Berge}
\begin{equation}\label{eq:5_1}
\rmi\dfrac{\partial\Psi}{\partial z}+\hat{\mathcal{D}}\Psi+|\Psi|^2\Psi-2\rmi|\Psi|^2\dfrac{\partial\Psi}{\partial\tilde\tau}+\hat{T}^{-1}\Delta_\perp \Psi=0 .
\end{equation}
Here, $k_z=-1/\omega$, $\tilde{\tau}=\tau-z/\omega_0^2$ is the time in the accompanying system of coordinates, 
$\omega_0=1$ is carrier frequency, $\hat{T}=1-\rmi\frac{\partial}{\partial\tilde{\tau}}$ is the operator correcting the approach of slowly changing amplitudes~\cite{Brabec}, and 
$$\hat{\mathcal{D}}=\sum\limits_{n\geq2}\frac{1}{\rmi^n n!} \left.\frac{\partial^nk_z}{\partial\omega^n}\right|_{\omega=\omega_0} \frac{\partial^n}{\partial\tau^n}.$$

For the field distributions $u(z,r_\perp ,\tau)$, which are localized in time and space, the following values are constant: 
\begin{subequations}\label{eq:6}
\begin{gather}
\iint ud\tau dr_\perp =0 , \label{eq:6a} \\ 
\mathcal{I}_\text{full}=\iint|u|^2d\tau dr_\perp  , \label{eq:6b} \\
\mathcal{H}_\text{full}=\iint\left[\Big|\int\limits_{-\infty}^{\tau} \nabla_\perp ud\tau'\Big|^2-\dfrac{|u|^4}{2}+\Big|\int\limits_{-\infty}^{\tau} ud\tau'\Big|^2\right]d\tau dr_\perp  . \label{eq:6c}
\end{gather}
\end{subequations}
Formula~\eqref{eq:6a} is responsible for the absence of the zeroth harmonic in the field distribution. The integral \eqref{eq:6b} determines the conservation of the ``number of quanta''. Formula~\eqref{eq:6c} is a Hamiltonian. Using the ``Hamiltonian'' nature of Eq.~\eqref{eq:5}, one can obtain the relationship
\begin{equation}\label{eq:7}
\mathcal{I}_\text{full} \dfrac{d^2\langle\rho_\perp^2\rangle}{dz^2}=8\mathcal{H}_\text{full}-8\iint\Big|\int_{-\infty}^{\tau}ud\tau'\Big|^2d\tau dr_\perp  ,
\end{equation}
which describes the variation of the efficient transverse width of the wave field in the process of self-focusing, 
\begin{equation}\label{eq:8}
\langle\rho^2_\perp \rangle = \frac{1}{\mathcal{I}_\text{full}} \iint r^2_\perp |u|^2d\tau dr_\perp   . 
\end{equation}
In what follows, we will consider the initial distributions of laser pulses with the negative Hamiltonian~$\mathcal{H}_\text{full}<0$, that are definitely collapsing in the transverse direction along a finite propagation path (see Eq.~\eqref{eq:7}).

In the case, where spatial effects are insignificant ($\Delta_\perp \equiv0$), there exists a class of stable soliton solutions~\cite{soliton}. Wave solitons of Eq.~\eqref{eq:5} can be represented by the two-parameter family of solutions having the form
\begin{equation}\label{eq:9}
u(z,\tau)=\sqrt{\gamma}G(\xi)\exp[i\omega_s(\tau+\gamma z)+i\phi(\xi)] ,
\end{equation}
where $\omega_s$ is the characteristic carrier frequency, 
$\gamma$ is the parameter determining the group velocity of the soliton, and 
$\xi=\omega_s(\tau-\gamma z)$. 
The soliton amplitude~$G(\xi)$ and the nonlinear phase~$\phi(\xi)$ obey the following equations: 
\begin{subequations}\label{eq:10}
\begin{gather}
\dfrac{d\phi}{d\xi}=\dfrac{G^2(3-2G^2)}{2(1-G^2)^2} , \label{eq:10a}\\
\int_{G_m}^G\dfrac{1-3G^2}{G\sqrt{\delta^2-F(G^2)}}dG=\pm(\xi-\xi_0) , \label{eq:10b}
\end{gather}
\end{subequations}
where $F(G^2)=G^2\left[3/2(1+\delta^2)-(4-5G^2)/4(1-G^2)^2\right]$, 
$G_m$ is the maximum amplitude of the soliton, and 
$\xi_0$ is the integration constant corresponding to the position of the maximum of the soliton amplitude. 
As seen from Eq.~\eqref{eq:10b}, solutions for the soliton amplitude~$G(\xi)$ depend only on the parameter~$\delta^2=1/(\omega_s^2\gamma)-1$ and exist at $0\leq\delta\leq\delta_\text{cr}\equiv\sqrt{1/8}$. The critical value gives the single-cycle soliton \cite{soliton}. An important feature of the wave solitons is a semi-bounded spectrum of their admissible solutions, \ie the presence of a boundary solution corresponding to the limiting soliton with the minimum possible pulse duration and, hence, the maximum possible amplitude. It should be noted that the existence of the limiting soliton is defined by the constraint $\int^{\infty}_{-\infty}ud\tau=0$, which is one of the integrals of Eq.~\eqref{eq:5}. At $\delta =\delta_\text{cr}$ the shortest duration is equal to $\tau_s^* =2.31\omega_s^{-1}$.

\begin{figure}[tpb]
	\centering
	\includegraphics[width = \linewidth]{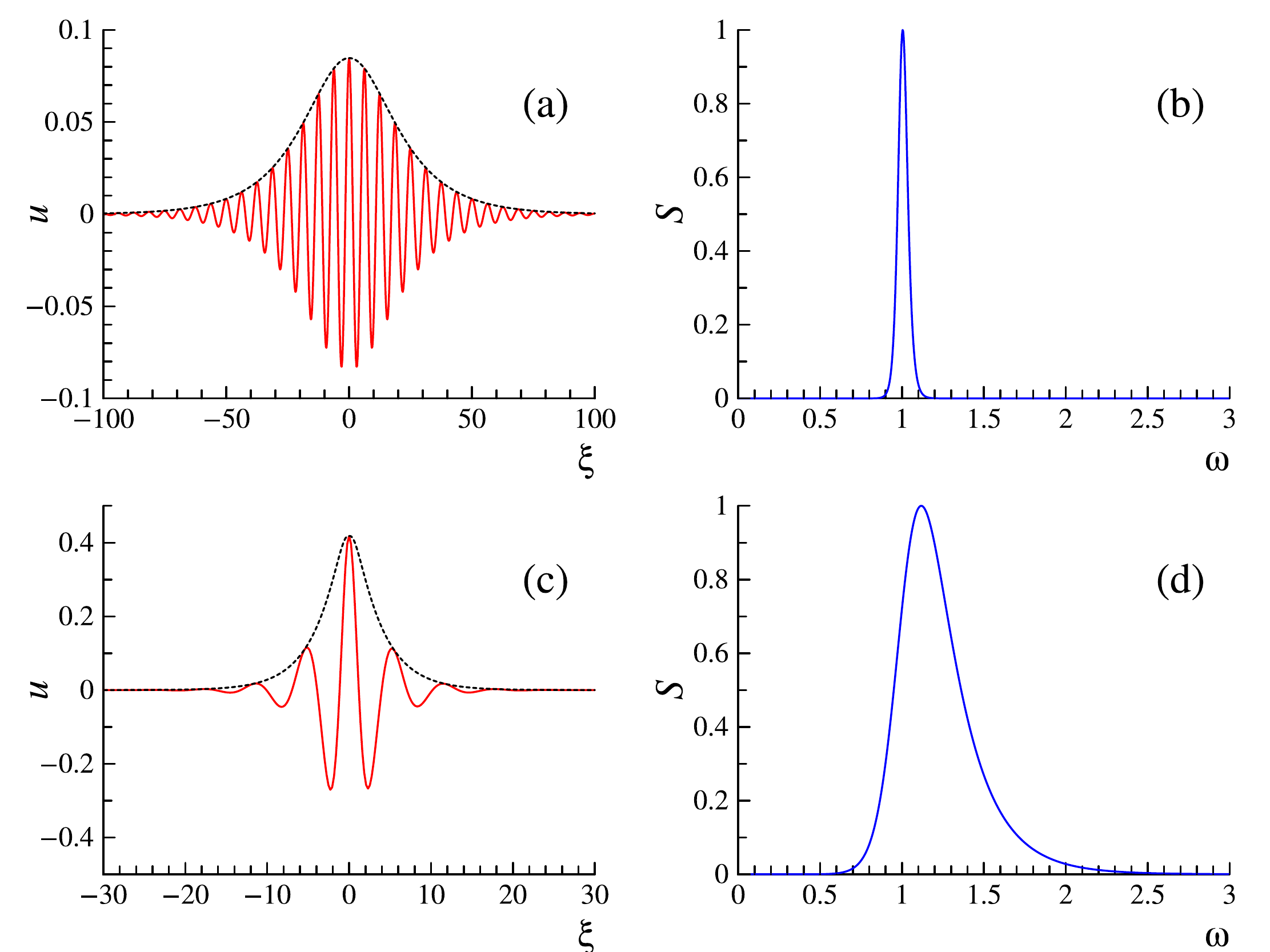}
	\caption{(Color online) 
		Exact soliton solutions for one of the field components of the field $u_x=Re(u)$ (red solid curves) corresponding to $\delta=0.06$, $\omega_s=1$ {\bf (a)} and $\delta=0.3$, $\omega_s=1$ {\bf (c)}. The dashed black line shows the distribution of the field envelope $\sqrt{\gamma}G(\xi)$. 
		Figures~{\bf (b)} and {\bf (d)} present distributions of spectral intensity for different $\delta$: {\bf (b)}  for $\delta=0.06$ and {\bf (d)} for $\delta=0.3$.}\label{ris:ris1}
\end{figure}

For the sake of comparison, Figs.~\ref{ris:ris1}{\bf (a,c)} present exact soliton solutions for two different values of the parameter~$\delta$:  {\bf (a)} for $\delta=0.06$ and {\bf (b)} for $\delta=0.3$. One can see that as the value of $\delta$ increases, the duration of the soliton decreases. Figures~\ref{ris:ris1}{\bf (b,d)} show the spectral intensities for various~$\delta$. It follows from the figure that as $\delta$ increases, the soliton spectrum becomes wide and asymmetric due to a strong frequency modulation in the pulse~\eqref{eq:10a}.

Now let us demonsrate the connection between the solution of Eq.~\eqref{eq:10b} with the well-known envelope solitons, which exist within the framework of NSE and its generalizations. Passing over to long quasi-monochromatic wave packets corresponds to the case of small values of the amplitudes~$G\ll1$. Keeping the terms of the order of $G^4$ in Eq.~\eqref{eq:10b}, we obtain the solitary solution for the envelope
\begin{equation}\label{eq:11}
G(\xi)=\dfrac{\sqrt2\delta}{\cosh(\delta\xi)}, \quad\phi_\xi\simeq0 .
\end{equation}
Typical distributions of the field and spectral intensity of the soliton at small $\delta=0.06$ are shown in Fig.~\ref{ris:ris1}{\bf (a,b)}. Keeping the terms of the next order of smallness, $G^6$, in Eq.~\eqref{eq:10b}, we obtain the NSE solution allowing for nonlinear dispersion~(amplitude dependence of the group velocity)
\begin{equation}\label{eq:12}
G(\xi)=\dfrac{2\delta}{\sqrt{1+\sqrt{1+12\delta^2}\cosh(2\delta\xi)}} , \quad \phi_\xi\simeq \dfrac32 G^2 .
\end{equation}
A distinctive feature of this solution is the presence of a sufficiently strong frequency modulation in a laser pulse. Typical distributions of the field and spectral intensity of the soliton at $\delta=0.3$ are shown in Fig.~\ref{ris:ris1}{\bf (c,d)}.

\section{Qualitative analysis of wave packet self-compression}\label{sec:3}

Now, let us turn to the problem of self-compression of wave packets in the self-focusing process. Equation~\eqref{eq:5} is much more complicated than the corresponding equation for the quasi-monochromatic radiation. Therefore, in order to obtain analytical relationships, first we turn to simplify Eq.~\eqref{eq:5_1}, where we neglect such important effects as high-order dispersion, nonlinear dispersion (amplitude dependence of the group velocity) and spatio-temporal self-focusing ($\hat{T}=1$)
\cite{litvak}
\begin{equation}\label{eq:13}
\rmi\dfrac{\partial\Psi}{\partial z}+\dfrac{\partial^2\Psi}{\partial\tau^2}+\Delta_\perp \Psi+|\Psi|^2\Psi=0 .
\end{equation}
Here, the second term describes anomalous media dispersion, the third term describes pulse diffraction, and the fourth term allows for the cubic nonlinearity of the medium. 

In this equation, the ``number of quanta'' and the Hamiltonian are also saved for localized distributions: 
\begin{subequations}\label{eq:14}
\begin{gather}
\mathcal{I}_q=\iint|\Psi|^2d\tau dr_\perp  , \label{eq:14a} \\
\mathcal{H}_q=\iint\left[\nabla_\perp \Psi|^2+|\partial_\tau \Psi|^2-\dfrac12|\Psi|^4 \right]d\tau dr_\perp   . \label{eq:14b}
\end{gather}
 \end{subequations}
Note that the issue of pulse self-focusing in media with anomalous and normal dispersion is addressed in many papers~\cite{Berge_Rasmussen,Petrova,Kosmatov,Kuznetsov,McLaughlin}. When a wave packet propagates in a nonlinear medium, it is affected simultaneously by dispersion and diffraction. At the same time, however, these two effects become interconnected due to the nonlinearity of the medium. This connection leads to the possibility of a spatio-temporal collapse. 

For equation \eqref{eq:13}, one can also write down a relationship for variation in the efficient scales (width and duration) of the wave packet (similar to Eq.~\eqref{eq:7}), which characterizes the global behavior of the system: 
\begin{subequations}\label{eq:15}
\begin{gather}
\mathcal{I}_{q}\dfrac{d^2\langle\rho^2_\perp \rangle}{dz^2}=8\mathcal{H}_{q}-8\iint|\partial_{\tau}\Psi|^2d\tau dr_\perp  , \label{eq:15a}\\
\mathcal{I}_{q}\dfrac{d^2\langle\rho_\perp ^2+\tau^2\rangle}{dz^2}=8\mathcal{H}_{q}-2\iint|\Psi|^4d\tau dr_\perp  , \label{eq:15b}
\end{gather}
\end{subequations}
where $\mathcal{I}_q \langle\rho_\perp ^2\rangle=\iint r_\perp ^2|\Psi|^2d\tau dr_\perp $ is the efficient transverse width of the wave packet, and 
$\mathcal{I}_q \langle\rho_\perp ^2+\tau^2\rangle=\iint(r_\perp ^2+\tau^2)|\Psi|^2d\tau dr_\perp $ is the total efficient scale of the wave packet. 
From here, as in the case of relationship~\eqref{eq:7} at the negative Hamiltonian~$\mathcal{H}_q<0$, one can make a certain conclusion about the collapse of not only efficient transverse width~\eqref{eq:15a}, but also the total  effective scale of the laser pulse~\eqref{eq:15b} in the case of its propagation in a nonlinear medium.

For a more detailed study of the evolution of the wave packet in a nonlinear medium, we turn to the {\em  variation approach}~\cite{Litvak_Fraiman}, which allows one to find an approximate solution of Eq.~\eqref{eq:13}. The essence of the method consists in finding solutions for this class of the functions~$\Psi=f(\vec{r};\sigma(z))$, where the set of parameters~$\sigma(z)$ depends on the evolution variable and is determined basing on the solutions of the corresponding system of differential equations. This system of equations is found on the basis of requirements of minimization of the action functional, in which the integration with respect to spatial variables~($r_\perp ,\tau$) is performed:
\begin{equation}\label{eq:16}
S[\Psi]=\iiint L(\Psi,\dot{\Psi},\nabla \Psi)dzd{\vec r}=\int\tilde{L}(\sigma, \dot{\sigma})dz
\end{equation}
As a result, the ``abbreviated'' Lagrangian 
\begin{equation}\label{eq:17}
\tilde{L}\equiv\int L\left(f(\vec{r};\sigma),\dot{\sigma_i}\dfrac{\partial f(\vec{r};\sigma)}{\partial\sigma_i},\nabla f(\vec{r};\sigma) \right)d{\vec r}
\end{equation}
yields a system of equations in a usual way: 
\begin{equation}\label{eq:18}
\dfrac{\partial\tilde{L}}{\partial\sigma_i}-\dfrac{d}{dz}\dfrac{\partial\tilde{L}}{\partial\dot{\sigma}_i}=0 .
\end{equation}
Note that the main difficulty of this method consists in the choice of the class of functions. If it is not well chosen, then the resulting approximate solution will be a little like a true solution of the original equation.

Let us use the variational approach in order to find the approximate solution of Eq.~\eqref{eq:13}. We will seek the solution in the class of Gaussian functions, \ie in the so-called aberration-free approximation
\begin{multline}\label{eq:19}
\Psi=\sqrt{\dfrac{\mathcal{I}_q}{\pi^{3/2}a_\perp ^2a_\parallel}}\exp\left(-\dfrac{(x^2+y^2)}{2a_\perp ^2}-\dfrac{\tau^2}{2a_\parallel^2}-\right.\\
-\left.\dfrac{\rmi\alpha(x^2+y^2)}{2} -\dfrac{\rmi\beta\tau^2}{2}\right) ,
\end{multline}
where $a_\perp (z)$, $a_\parallel(z)$, $\alpha(z)$ and $\beta(z)$ are the parameters of the function, which depend on the evolution variable~$z$. The advantages of the Gaussian profile of solution~\eqref{eq:19} consist in its well-localization and in the absence of singularity at center.

Substituting expression Eq.~\eqref{eq:19} to the action functional \eqref{eq:16} with Lagrangian
\begin{multline}\label{eq:20}
L=\left[\dfrac{\rmi}{2}\left(\Psi^{\star}\partial_z\Psi-\Psi\partial_z\Psi^{\star}\right)-\right.\\
\left.-\left|\partial_\tau\Psi\right|^2-\left|\nabla_\perp \Psi\right|^2+\dfrac{1}{2}|\Psi|^4\right],
\end{multline}
we obtain an ``abbreviated'' Lagrangian 
$\tilde{L}$
\begin{multline}\label{eq:21}
\tilde{L}=\dfrac{\mathcal{I}_q}{4}\left(2a_\perp ^2\dfrac{d\alpha}{dz}+a_\parallel^2\dfrac{d\beta}{dz}\right)-\dfrac{\mathcal{I}_q}{2a_\parallel^2}\left(1+\beta^2a_\parallel^4\right)-\\-\dfrac{\mathcal{I}_q}{a_\perp ^2}\left(1+\alpha^2a_\perp ^4\right)+\dfrac{\mathcal{I}_q^2}{4\sqrt{2}\pi^{3/2}a_\perp ^2a_\parallel} .
\end{multline}
Variation with respect to the variables $\sigma=\lbrace a_\perp ,a_\parallel,\alpha,\beta\rbrace$ yields the following equations
\begin{subequations}
\begin{itemize}
 \item from variation with respect to $\alpha$: 
\begin{equation}\label{eq:23}
\alpha=-\dfrac{1}{2a_\perp }\dfrac{da_\perp }{dz}
\end{equation}
\item from variation with respect to  $\beta$: 
\begin{equation}\label{eq:24}
\beta=-\dfrac{1}{2a_\parallel}\dfrac{da_\parallel}{dz}
\end{equation}
\item from variation with respect to $a_\perp $: 
\begin{equation}\label{eq:25}
\dfrac{d^2a_\perp }{dz^2}=\dfrac4{a_\perp ^3}-\dfrac{\mathcal{I}_q}{\sqrt{2}\pi^{3/2}a_\perp ^3a_\parallel}
\end{equation}
\item from variation with respect to $a_\parallel$: 
\begin{equation}\label{eq:26}
\dfrac{d^2a_\parallel}{dz^2}=\dfrac4{a_\parallel^3}-\dfrac{\mathcal{I}_q}{\sqrt{2}\pi^{3/2}a_\perp ^2a_\parallel^2} .
\end{equation}
\end{itemize}
\end{subequations}
It is seen from Eqs.~\eqref{eq:25}-\eqref{eq:26} that the decrease in width and duration of the laser pulse in the process of radiation self-focusing will take place in the case, where the initial packet duration $a_\parallel(0)$ satisfies the following relationship:
\begin{equation}\label{eq:27}
\dfrac{4\sqrt2\pi^{3/2}a_\perp ^2(0)}{\mathcal{I}_q}\ll a_\parallel(0)\ll\dfrac{\mathcal{I}_q}{4\sqrt2\pi^{3/2}} .
\end{equation}
In the case of a great difference in the scales~($a_\perp \ll a_\parallel$), as follows from Eqs.~\eqref{eq:25}-\eqref{eq:26}, the compression will occur only along the transverse coordinate. In this case, when the initial width and duration are not strongly different, the scales along two coordinates will line up in the process of nonlinear dynamics~\cite{Berge_Rasmussen}. The regime of spherically symmetric collapse, when the longitudinal and transverse scales are equal ($a_\perp =a_\parallel$), was studied in~\cite{book}.

Now, we will consider the most interesting special case of the self-action regime, where the pulse envelope have longitudinal scale being much less than the transverse ones ($a_\parallel\ll a_\perp $).
In this case, equation~\eqref{eq:26} is an equation with small coefficient at highest derivative. Its ``slow'' motion is
\begin{equation}\label{eq:29b}
 a_\parallel=\dfrac{4\sqrt2\pi^{3/2}a_\perp ^2}{\mathcal{I}_q}.
\end{equation}
At ``slow'' motion the soliton-like law for pulse duration is fulfilled: $d^2 a_\parallel/dz^2 \approx 0$. This corresponds to
an adiabatic decrease in the duration of the soliton-like wave packet. Substituting Eq.~\eqref{eq:29b} into Eq.~\eqref{eq:25} we obtain:
\begin{equation}\label{eq:29a}
 \dfrac{d^2a_\perp }{dz^2}=\dfrac4{a_\perp ^3}-\dfrac{\mathcal{I}_q^2}{8\pi^3a_\perp ^5}.
\end{equation}
It follows from Eq.~\eqref{eq:29b} that the pulse duration decreases in proportion to square of pulse width, \ie it is inversely proportional with the field intensity at the pulse axis. Note that for oblate pulses ($a_\parallel\ll a_\perp $), the self-focusing condition is fulfilled automatically (the right-hand part of Eq.~\eqref{eq:29a} should be negative), since the second term in Eq.~\eqref{eq:29a}, which is responsible for the medium nonlinearity, exceeds the first term describing the pulse diffraction, 
\begin{equation}\label{eq:30}
\dfrac{a_\perp }{a_\parallel}=\dfrac{\mathcal{I}_q}{4\sqrt2\pi^{3/2}a_\perp }\gg1,~\dfrac{\mathcal{I}_q^2}{8\pi^3a_\perp ^5}=\dfrac{4}{a_\perp ^3}\left(\dfrac{a_\perp }{a_\parallel}\right)^2\gg\dfrac{4}{a_\perp ^3} .
\end{equation}

In what follows, we will neglect the first term in Eq.~\eqref{eq:29a} for such distributions. This allows us to find the variation laws for the longitudinal and transverse pulse scales depending on $z$. The solution of Eq.~\eqref{eq:29a} allowing for Eq.~\eqref{eq:30} can be represented in quadratures under the following initial conditions at $z=0$: $\dot{a}_\perp=0$, $a_\perp=a_{\perp0}$
\begin{equation}\label{eq:31}
 \dfrac{da_\perp }{dz}=-\dfrac{\mathcal{I}_q}{4\pi^{3/2}}\sqrt{\dfrac1{a_\perp ^4}-\dfrac1{a_{\perp0}^4}} .
\end{equation}
To solve this equation, we will use the fact that as the pulse width decreases ($a_\perp <a_{\perp0}$), the second term in Eq.~\eqref{eq:31} becomes less than the first one and, correspondingly, it can be neglected. As a result, we obtain approximate solutions for the pulse width $a_\perp$ and duration~$a_\parallel$
\begin{subequations}\label{eq:32}
\begin{gather}
a_\perp (z)\simeq a_{\perp0}\left(1-\dfrac{3\mathcal{I}_qz}{4\pi^{3/2}a_{\perp0}^3} \right)^{1/3} , \label{eq:32a} \\
a_\parallel\simeq a_{\parallel0}\left(1-\dfrac{3\mathcal{I}_qz}{4\pi^{3/2}a_{\perp0}^3} \right)^{2/3} , \label{eq:32b}
\end{gather}
\end{subequations}
where $a_{\perp0}$ is the initial pulse width, and $a_{\parallel0}$ is the initial pulse duration. 
It should be noted that, as follows from Eqs.~\eqref{eq:32}, the pulse duration $a_\parallel$ always stays smaller than the pulse width $a_\perp $ ($a_\perp \gg a_\parallel$), \ie the anisotropy of the wave packet distribution is retained. 

Using formulas~\eqref{eq:29b} and \eqref{eq:32b}, one can evaluate the compression length~$z_\text{comp}$, at which the pulse duration will turn to zero, [$a_\parallel(z_\text{comp})=0$]: 
\begin{equation}\label{eq:33}
z_\text{comp} =\frac{4\pi^{3/2} a_{\perp0}^3}{3 \mathcal{I}_q} =\dfrac{a_{\perp0}a_{\parallel0}}{3\sqrt2}.
\end{equation}
Let us rewrite this formula using dimensional units, 
\begin{equation}\label{eq:34}
z_\text{comp}=\dfrac{\sqrt2}{3}\dfrac{\omega_0}{\omega_D} \dfrac{\omega_0 a_{\parallel0}}{c} a_{\perp0},
\end{equation}
It is seen from Eq.~\eqref{eq:34} that the length of the medium, at which the duration of the soliton turns to zero, is proportional to initial duration and width of the wave packet and decreases with decreasing of media linear dispersion. 

\section{Ultimate possibilities of laser pulse self-compression in the process of radiation self-focusing}\label{sec:4}

The qualitative analysis performed in the previous section on the basis of the NSE equation~\eqref{eq:13} shows that the pulse duration will decrease adiabatically down to the zero in the self-focusing process. The question about a minimal pulse duration at self-compression arises. Evidently, as the duration of the wave packet decreases, additional effects start to show themselves, which can limit the shortening of the pulse duration, specifically, the dependence of the group velocity of the packet on the field amplitude $\rmi|\Psi|^2\partial_\tau\Psi$, dispersion of the group velocity of a higher order $\sum_{n>2} (1/\rmi^n n!) (\partial^nk_z/\partial\omega^n) \partial^n\Psi/\partial\tau^n$, where $k_z$ is the wave number, and the spatio-temporal self-focusing (see Eq.~\eqref{eq:13}). In this case, we should return to initial equation~\eqref{eq:5} for analysis of the ultimate self-compression of laser pulses in the process of radiation self-focusing. As it has been already noted, the initial distributions of the wave field at~$\mathcal{H}_\text{full}<0$ will undergo self-focusing in the transverse direction, since collapse condition~\eqref{eq:7} is fulfilled for such distributions. 

For qualitative studies of the ultimate self-compression of laser pulses in the framework of initial equation~\eqref{eq:5}, it is convenient to pass over from the laboratory system of coordinates to the system of coordinates, which collapses towards a certain point~($r_\perp =0$, $z=z_0$). Let us represent the field $u(z,r_\perp ,\tau)$ as 
\begin{equation}\label{eq:35}
u=\dfrac{\digamma\left(\zeta ,\eta ,\theta \right )}{\rho(z)} ,
\end{equation} 
where the new variables are 
\begin{equation}\label{eq:36}
\zeta=\int \dfrac{dz}{\rho^2(z)} , \quad\eta = \dfrac{r_\perp }{\rho(z)} , ~\theta =\tau -\dfrac{\rho_z}{4\rho}r_\perp ^2\ .
\end{equation}
Here, $\zeta$ is the new reference scale for the evolution variable, for which the moment of singularity formation is shifted to infinity. The function~$\rho(z)$ describes the variation in the transverse width of the field. This representation of the solution of Eq.~\eqref{eq:5} allows for two processes simultaneously, namely, radiation self-focusing and formation of a characteristic ``horsehoe'' structure of the field distribution, which is determined by the variable~$\theta$.

As a result of applying transformations \eqref{eq:35} and \eqref{eq:36} to equation~\eqref{eq:5}, we arrive at the following equation:
\begin{equation}\label{eq:37}
\dfrac{\partial}{\partial\theta}\left(\dfrac{\partial\digamma}{\partial\zeta}+\dfrac{\partial }{\partial\theta}\left(|\digamma|^2\digamma\right)-\dfrac{\rho_{zz}\rho^3}{4}\eta^2\dfrac{\partial\digamma}{\partial\theta} \right)+
\rho^2(z)\digamma=\Delta_{\eta}\digamma . 
\end{equation}
Transformation to the ``collapsing'' system of coordinates allows us, as in the case of quasimonochromatic radiation~(Eq.~\eqref{eq:13}), to segregate the self-focusing process in the system and reduce the problem to studying the quasi-one-dimensional longitudinal evolution of the pulse with respect to the variable~$\theta$. The characteristic transverse scale of the quasi-waveguide structure in new variables is of the order of unity. 

Now, we represent the field in the near-axis region ($\eta\approx0$) as $\digamma=\mathcal{A}(\zeta,\theta)\cdot\left(1-\eta^2/4\right)$. Substituting this formula to Eq.~\eqref{eq:37} and setting the coefficients in front of $\eta^0$ and $\eta^2$ equal to zero, we find equations for $\mathcal{A}$ and $\rho$:
\begin{subequations}\label{eq:38}
\begin{gather}
\dfrac{\partial^2\mathcal{A}}{\partial\zeta\partial\theta}+\rho^2\mathcal{A}+\dfrac{\partial^2}{\partial\theta^2}(|\mathcal{A}|^2\mathcal{A})\simeq0 , \label{eq:38a} \\
\dfrac{d^2\rho}{dz^2}\simeq-2\dfrac{\overline{|\mathcal{A}|^4}}{\rho^3} . \label{eq:38b}
\end{gather}
\end{subequations}
When obtaining Eqs.~\eqref{eq:38b}, we averaged $|\mathcal{A}|^4$ with respect to the pulse shape \Big($\overline{|\mathcal{A}|^4}=\frac1{\mathcal{I}_\text{full}}\int|\mathcal{A}|^4 d\theta$\Big), since the characteristic scale of the field $\rho(z)$ is a function of $z$ only, by assumption. Formula~\eqref{eq:38a} is valid, when the threshold for self-focusing is exceeded significantly, which is valid for oblate wave packets, as we have mentioned in the previous section. 

One can see that Eq.~\eqref{eq:38a}, which describes the field dynamics in the near-axis region~($\eta\approx0$), coincides with Eq.~\eqref{eq:5} at $\Delta_\perp \equiv0$ (in the absence of spatial effects). The second term in Eq.~\eqref{eq:38a} describes weakening of the medium dispersion role with the decreasing of the pulse width $\rho(z)$. As it was mentioned at the end of Section~\ref{sec:2}, within the framework of Eq.~\eqref{eq:38a} at a constant value of $\rho$, there is a family of soliton solutions~(\ref{eq:9}-\ref{eq:10}) having the shape, which is similar to that of Schr\"odinger soliton~\eqref{eq:11}. An important difference of these solutions is the presence of a strong frequency modulation \eqref{eq:12} in solitons at short durations.

Let's suppose now that the function $\rho(\zeta)$ varies smoothly  in new variables
$$\frac{1}{\rho} \frac{d\rho}{d\zeta} \ll \Delta \omega |\mathcal{A}_\text{max}|^2, $$
where $\Delta\omega$ is the spectral width of the laser pulse. So, the soliton parameters will be adjusted smoothly in the self-focusing process and an adiabatic increase in the soliton amplitude $u\propto {\mathcal A}/{\rho}$ will take place as the pulse width $\rho(z)$ will decrease~Eq.~\eqref{eq:38b}. The solution of Eq.~\eqref{eq:38a} can be written for $\mathcal{A}(\zeta,\theta)=\mathcal{B}\exp(\rmi\phi)$ in the following form depending on the current soliton duration $\tau \propto 1/{\delta}$:
\begin{equation}\label{eq:39}
\mathcal{B}(\zeta,\theta)\simeq\dfrac{\sqrt{2}\rho\delta}{\cosh(\delta\theta)}, \quad \dfrac{d\phi}{d\theta} \simeq \begin{cases}
0, & \textrm{at } \delta \ll 1 \\
\dfrac32\mathcal{B}^2, & \textrm{at } \delta\gtrsim 0.1
\end{cases},
\end{equation}
Allowing for preservation of the total ``number of quanta'',
\begin{equation}\label{eq:40}
\mathcal{I}_\text{full}=4\pi\delta\rho^2 
\end{equation}
it is convenient to rewrite the solution of Eq.~\eqref{eq:39} using this quantity. Thus, we manage to get rid of the parameter $\delta$, which is now related via the energy $\mathcal{I}_\text{full}$ in the laser pulse and the current pulse width~$\rho(z)$. As a result, we obtain the final system of equations for the dynamics of the wave packet: 
\begin{subequations}\label{eq:41}
\begin{gather}
\mathcal{B}(\zeta,\theta)\simeq\dfrac{\mathcal{I}_\text{full}}{2^{3/2}\pi\rho} \dfrac{1}{\cosh\left(\dfrac{\mathcal{I}_\text{full}}{4\pi\rho^2}\theta\right)}  , \label{eq:41a} \\
\dfrac{d^2\rho}{dz^2}\simeq -\dfrac{\alpha \mathcal{I}_\text{full}^2}{8\pi^3\rho^5} , \label{eq:41b}
\end{gather}
\end{subequations}
where $\alpha$ is a number of the order of unity. In this case, the frequency modulation in the soliton~$\phi(\theta)$ depending on the current duration of the soliton
\begin{equation}
\tau_p\propto \dfrac{4\pi\rho^2}{\mathcal{I}_\text{full}}
\end{equation}
obeys formula~\eqref{eq:39}.

One can see from formula~\eqref{eq:41a} that as the transverse pulse width~$\rho$ decreases, the soliton duration~$\tau_p$ decreases in the near-axis region ($\eta\approx0$). Thus, due to the process of pulse self-focusing, the soliton duration will decrease adiabatically. Additionally, the soliton velocity $1/{\gamma} \simeq {\omega^2}/{\rho^2}$ will tend to the velocity of light $c/{\sqrt{\varepsilon_0}}$. One can see that for the pulse width, Eq.~\eqref{eq:41b} coincides up to numerical coefficient with Eq.~\eqref{eq:29a}.

Here, if a Schr\"odinger-like soliton with $\delta \ll 1$ and, correspondingly, without a frequency modulation $\phi(\theta)\simeq0$ is sent to the entrance of the nonlinear medium, then, as the pulse duration decreases, a strong frequency modulation $\phi(\theta)\simeq\frac32 \int_{-\infty}^{\theta}|\mathcal{A}(\theta')|^2d\theta'$ will arise in the pulse, which will manifest itself in the wave packet spectrum~(see Fig.~.~\ref{ris:ris1}{\bf (d)}).

As it has been noted in section \ref{sec:3}, a distinctive feature of considered wave solitons (within the framework of Eq.~\eqref{eq:5} at $\Delta_\perp \equiv0$) is the semi-bounded spectrum of their admissible solution, \ie the presence of the bounding solution $0\leq\delta\leq\delta_\text{cr}=\sqrt{1/8}$. Therefore, the maximum degree of self-compression will be determined by this bounding soliton, whose duration is comparable with the optical cycle. 
This is the difference between the final regime of laser pulse self-compression and the regime, which we considered on the basis of nonlinear Schr\"odinger equation~\eqref{eq:13}.

\section{Study of the stability of the 3D wave packet relative to the filamentation instability}\label{sec:5}

The considered key idea of the laser pulse compression is an adiabatic decrease of the soliton duration in the self-focusing process. The initial amplitude of the soliton is related to its duration by the following formula:
\begin{equation}\label{eq:52}
u_\text{in}=\sqrt{2\gamma}\delta_0=\sqrt{\dfrac{\gamma}{2}}\dfrac1{\tau_p^{in}} ,
\end{equation}
where $\gamma$ is the soliton velocity, $\tau_p^{in}=1/(2\delta_0)$ is the initial duration, and 
$\delta_0$ is the initial parameter of the soliton. The total energy in the spatio-temporal bounded quasi-soliton distribution of the field is determined by formula~\eqref{eq:40}. For practical realization of the proposed quasi-soliton method of pulse self-compression, of interest are wide-aperture wave packets, in which the energy flow $\mathcal{I}_s=\int |u|^2d\tau$ is completely determined by the nonlinearity and dispersion of the medium. In this connection, one faces an important problem of studying the stability of the considered regime of wave packet compression in relation to various spatio-temporal perturbations of the initial structure. 

The conclusion about the character of filamentation instability of continuous radiation is usually made on the basis of analysis of the instability of a plane wave~\cite{Bespalov_Talanov}. The transverse perturbations of the wave front in the interval~$0<\kappa_\perp \leq\kappa_\text{cr}=\sqrt{2}\mathcal{B}_0$ increase exponentially $\propto \exp(\Gamma z)$, as the wave packet propagates along $z$. Here the growth rate $\Gamma^2=\kappa_\perp ^2(2\mathcal{B}_0^2-\kappa_\perp ^2)$, where $\mathcal{B}_0$ is the amplitude of the plane wave. The growth rate~$\Gamma$ has the maximum value~$\Gamma_\text{max}=\mathcal{B}_0^2$ at $\kappa_{\perp m}=\mathcal{B}_0$. In the two-dimensional case, this indicates that the wave field splits into a set of beams with a power being of the order of magnitude of the critical value for self-focusing. This conclusion is also valid for inhomogeneous wave structures with the soliton distribution along one of the coordinates \cite{Zakharov}. A more general consideration on the basis of the NSE allowing for time dispersion also leads to the existence of an instability with the growth rate~\cite{litvak}
\begin{equation}\label{eq:56}
\Gamma=\pm\sqrt{(\Omega^2+\kappa_\perp ^2)(2\Psi_0^2-\Omega^2-\kappa_\perp ^2)} .
\end{equation}
In the case of the modulation frequency~$\Omega$ being equal to zero ($\Omega=0$), formula~\eqref{eq:56} describes instability of the plane wave discussed above, where we talked about the consequences of wave evolution. For $\kappa_\perp =0$, the corresponding instability is known as the modulation instability. Generally, the instability has the spatio-temporal character. 

However, in the case of a shorter soliton pulse and, therefore, a more intense peak power, additional nonlinear effects, such as the dependence of the group velocity on intensity, should be taken into account, which in the first approximation modifies the NSE equation into the so-called derivative nonlinear Schr\"odinger equation \eqref{eq:5_1}.

To analyze the stability of the plane wave relative to the perturbations, we will seek for the solution in the form 
\begin{equation}\label{eq:58}
\Psi(z,\tau,r_\perp )=[u_0+v(z,\tau,r_\perp )]e^{\rmi\phi(z,\tau)} , |v|\ll u_0 ,
\end{equation}
where $u_0$ is the amplitude of the incident wave, 
Substituting formula~\eqref{eq:58} to Eq.~\eqref{eq:5_1}, in the zeroth approximation with respect to $v$ we obtain that $\phi=u_0^2z+2u_0^2\tau$.  As a result, the linearized equation of the first order of smallness with respect to $v$ will acquire the form: 
\begin{equation}\label{eq:59}
\rmi\dfrac{\partial v}{\partial z}+\hat{\mathcal{D}} v + \hat{T}^{-1}\Delta_\perp v+u_0^2(v+v^{\star})+2\rmi u_0^2\dfrac{\partial v}{\partial\tau}=0.
\end{equation}

This equation differs from the usual equation for analysis of stability of the plane wave \cite{Bespalov_Talanov} and from more general equation with high-order dispersion terms \cite{Berge} by the presence of the last term and is connected with the allowance for the nonlinear medium dispersion. We will seek for the solution of the equation in the form $v=a+\rmi b$, where $a,b\propto e^{\Gamma z+\rmi\Omega\tau-\rmi\kappa_\perp r_\perp }$. As a result, we obtain a system of two homogeneous equations, which has a non-trivial solution in the case, where $\Omega$ and $\kappa_\perp $ satisfy the following dispersion relation: 
\begin{multline}\label{eq:60}
\Gamma=-2\rmi u_0^2 \Omega\pm \\ \pm \sqrt{\left(K(\Omega)+\frac{\kappa_\perp ^2}{1+\Omega} \right) \left(2u_0^2 - K(\Omega) - \frac{\kappa_\perp ^2}{1+\Omega} \right)} ,
\end{multline}
where $K(\Omega) = k_z(\omega_0+\Omega)-k_z(\omega_0)-\Omega/v_\text{gr}$.
As a result, we find that a drift of perturbations along the pulse takes place at a rate determined by the imaginary part of~$\Gamma$ in the pulse frame of reference, $\tau=\omega_0(t+z/v_\text{gr})$. Indeed, in the structure of perturbations~$a, b\propto e^{\Gamma(\Omega,\kappa_\perp ) + \rmi\Omega\tau - \rmi\kappa_\perp r_\perp }$ along with the exponential increase in the perturbations at $0\leq\sqrt{K(\Omega)+\kappa_\perp ^2/(1+\Omega)}\leq\sqrt{2}u_0$ (which is described by the second term in formula~\eqref{eq:60}). This is easily seen, when one considers the time dependence of $v$: 
\begin{equation}\label{eq:61}
v\propto\int S(\Omega) \rme^{\Gamma z} \exp(-2\rmi u_0^2\Omega z-\rmi\Omega\tau)d\Omega\propto f(\tau+2u_0^2z) ,
\end{equation}
where $S(\Omega)$ is the perturbation spectrum. It is seen from Eq.~\eqref{eq:61} that the perturbations move with the relative velocity $2u_0^2$. 
Consequently, the homogeneous solution persists to be unstable. However, the instability changes its type and becomes \emph{convective} with group velocity $\partial \Imag \Gamma/\partial \Omega = -2u_0^2$. Hence, for laser pulses with a duration less than a certain value $\tau_p<\tau_\text{cr}$, filamentation instability has no time to develop. 

Next, we estimate the critical pulse duration $\tau_\text{cr}$ at which the perturbation growth is stabilized due to the drift to the rear pulse front where its amplification becomes negligible. 
Dangerous perturbations slip down to the rear side of the pulse, because their group velocity is less than the soliton one. 
The ``slipping''  length $z=z_*$ at which the perturbation $v$ shifts by half the pulse duration is determined as $2 u_0^2 z_* \simeq \tau_{p}/2$. This length should be smaller than $\ln (\delta \Psi/u_0) \approx 15 \ldots 20$ of maximal perturbation growth length $1/\Gamma_\text{max} = 1/u_0^2$. As the result, we obtain the following inequality on the critical length (in dimensional units):
\begin{equation}\label{eq:63}
\tau_p \lesssim \tau_\text{cr}=20\pi/\omega.
\end{equation}
In other words, if the pulse duration is less than about ten optical cycles then the transverse modulation instability is suppressed. Moreover, this suppression doesn't depend on the pulse amplitude or media properties.

\begin{figure}[htpb]
	\centering
	\includegraphics[width = \linewidth]{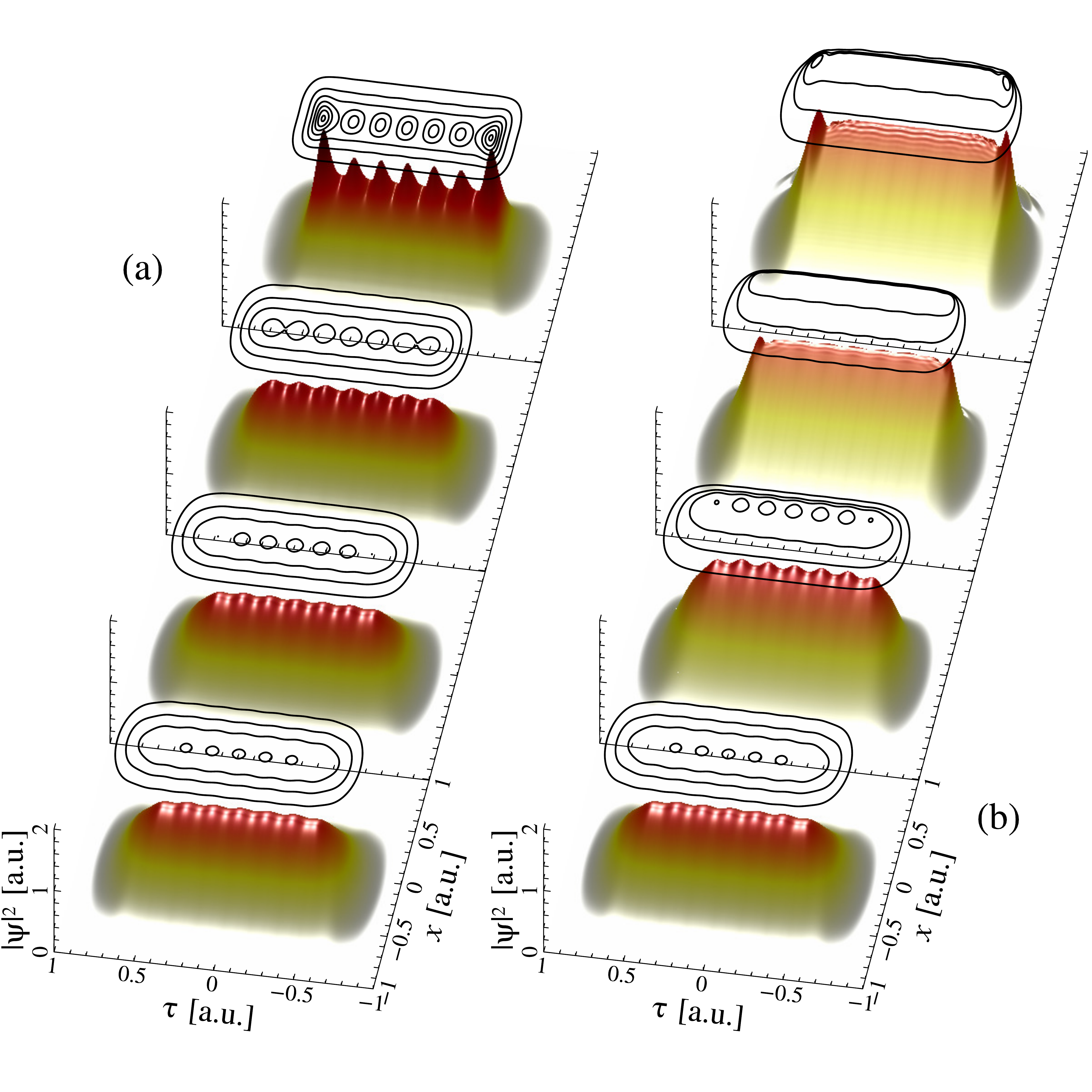} 
	\caption{(Color online)  Dynamics of wave packet intensity $|\Psi(z,x,\tau)|^2$ in Eq.~\eqref{eq:5_1} for two cases: {\bf (a)} in the absence of nonlinear dispersion; {\bf (b)} with nonlinear dispersion taken into account. The black curves show level lines. }\label{ris:ris8}
\end{figure}

Let us consider the results of numerical simulations using Eq.~\eqref{eq:5_1} (\ie keeping lowest nonzero term in $\hat{\mathcal{D}}$) to confirm the qualitative analysis. Since we study the problem of structural stability, we restrict numerical simulation by the case (2D+1) to simplify the analysis, because the result depends weakly on dimension of the problem. Figure~\ref{ris:ris8} shows the results of numerical simulation for the wave packet with input distribution
\begin{equation}\label{eq:61_1}
\Psi(x,\tau)=\Psi_0\exp\left[-\dfrac12\left( \big(\frac{\tau}{\tau_p}\big)^2 + \big(\dfrac{x}{a_x}\big)^8 \right) \right]
\end{equation}
at two different cases demonstrating different evolution of the filamentation instability. The evolution of pulse intensity $|\Psi(z,x,\tau)|^2$ in the absence of the fourth term in Eq. \eqref{eq:5_1}, \ie by neglecting pulse steepening, is shown in Fig. \ref{ris:ris8}{\bf (a)}. The pulse evolution with nonlinear dispersion taken into account is shown for comparison in Fig. \ref{ris:ris8}{\bf (b)}. As seen in Fig. \ref{ris:ris8}{\bf (a)}, filamentation instability splits  the wave packet into separate pulses in the transverse direction \cite{Bespalov_Talanov,Zakharov}. Here, the contour level lines are shown by the green color. However, if nonlinear dispersion is taken into account in Eq. \eqref{eq:5_1}, the instability is stabilized. It is clear from Fig. \ref{ris:ris8}{\bf (b)} that the inhomogeneities pass to the rear part of the pulse and cease to grow. This is the key difference between the considered pulse evolution mode and the laser pulse evolution within the framework of an ordinary nonlinear Schr\"odinger equation.

\begin{figure*}[htpb]
	\centering
	\includegraphics[width = \linewidth]{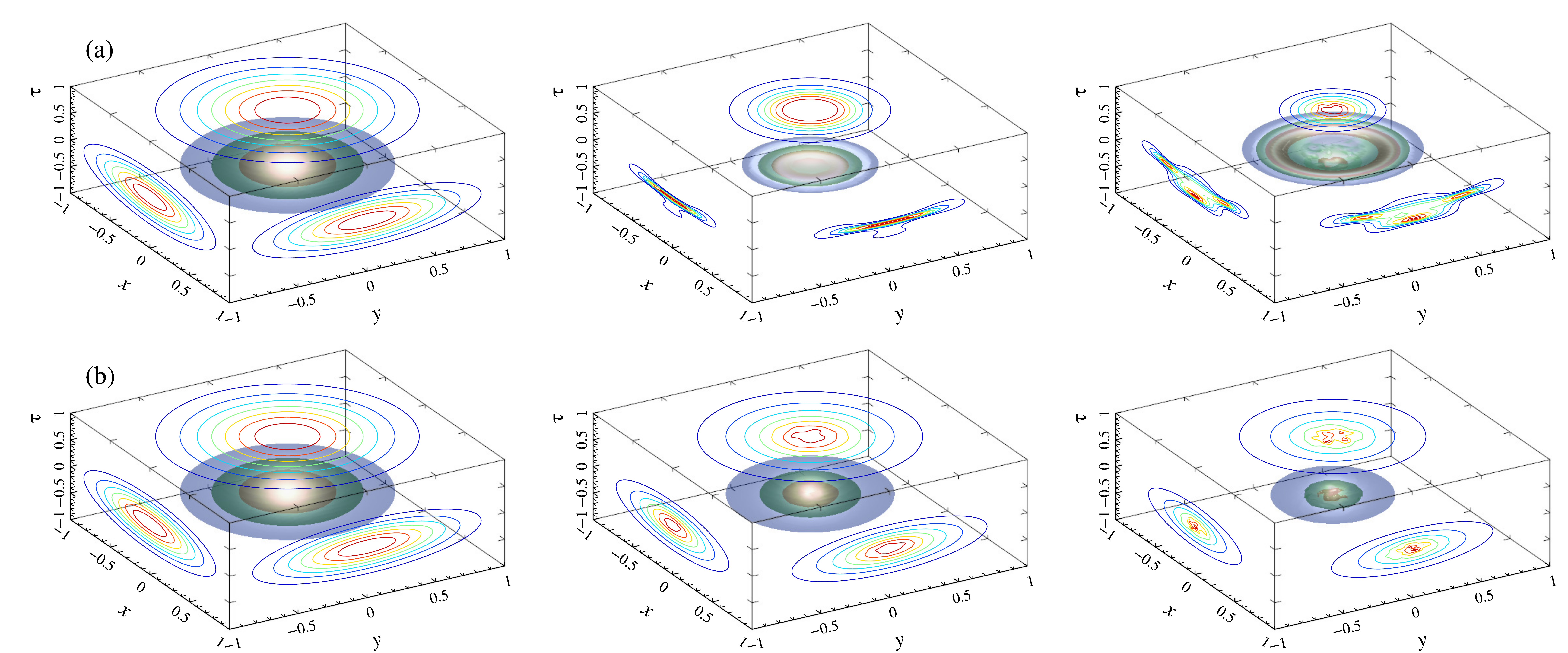}
	\caption{(Color online) 
		Dynamics of the circularly polarized soliton~$|u(z,x,y,\tau)|$ within the framework of Eq.~\eqref{eq:5} for two different initial durations in the process of self-focusing: {\bf (a)} $\tau_p=10\cdot T_0$, {\bf (b)} $\tau_p=30\cdot T_0$, where $T_0$ is the optical cycle. 
	}\label{ris:ris3}
\end{figure*}

A similar process of the filament instability stabilizing for few-cycle pulses occurs also for the laser pulse dynamics in the framework of the original wave equation \eqref{eq:5} for (3D+1) case. Figure~\ref{ris:ris3} shows the results of evolution of two different Gaussian laser pulses comprising 30 {\bf (a)} and 10 {\bf (b)} optical cycles with initial noise level of about $10^{-4}$ of pulse amplitude. The process of pulse self-focusing demonstrated in the $(x,y)$~plane occurs in both cases together with the strong pulse compression shown in the $(y,\tau)$~plane. However, the longer pulse compression is accompanied by simultaneous development of the filamentation instability (see the $(x,y)$ cross section in Fig.~\ref{ris:ris3}(a)), \ie the pulse splits into separate filaments. For the shorter pulse (Fig.~\ref{ris:ris3}(b)) the spatial structure of the pulse remains smooth in the process of adiabatic shortening of the pulse duration (see the $(x,y)$~cross section). So, the transverse instability can be suppressed, and no violation of the pulse symmetry is observed. Correspondingly, we perform a thorough numerical study of the pulse dynamics for the axisymmetric case.

Thus, as shown by the results of the analytical and numerical study, laser pulses with durations of less than ten optical cycles are not a subject to filamentation instability due to medium nonlinear dispersion. 

\section{Results of numerical simulation}\label{sec:6}
It was shown in the previous section that the spatial modulation instability can be suppressed, and, correspondingly, the pulse symmetry will not be violated. To perform detailed numerical analysis of soliton self-compression on the basis of Eq.~\eqref{eq:5}, in what follows we will turn to studying the dynamics of axisymmetric circular polarized laser pulses. 

As the initial distribution of the laser pulse, we will specify the soliton-like distribution (Eq.~\eqref{eq:10}) along $\tau$ and the transverse Gaussian distribution with the characteristic scale $a$:
\begin{equation}\label{eq:64}
u=\mathcal{N}\sqrt{\gamma}G(\tau)\exp\left(i\tau+i\phi(\tau)-\dfrac{r^2}{2a^2}\right) .
\end{equation}
Here, $\gamma$ is the soliton velocity, and $G$ and $\phi$ are the amplitude and phase distribution of the soliton, which are found by solving system of equations~\eqref{eq:10}, and the parameter~$\mathcal{N}$ characterizes the number of solitons, into which the initial laser pulse will be decomposed at the asymptotic stage within the framework of the one-dimensional problem~($\Delta_\perp \equiv0$).

\subsection{Single-soliton dynamics ($\mathcal{N}\simeq1$)}\label{sec:sec6_1}

\begin{figure}[htpb]
	\center{\includegraphics[width=\linewidth]{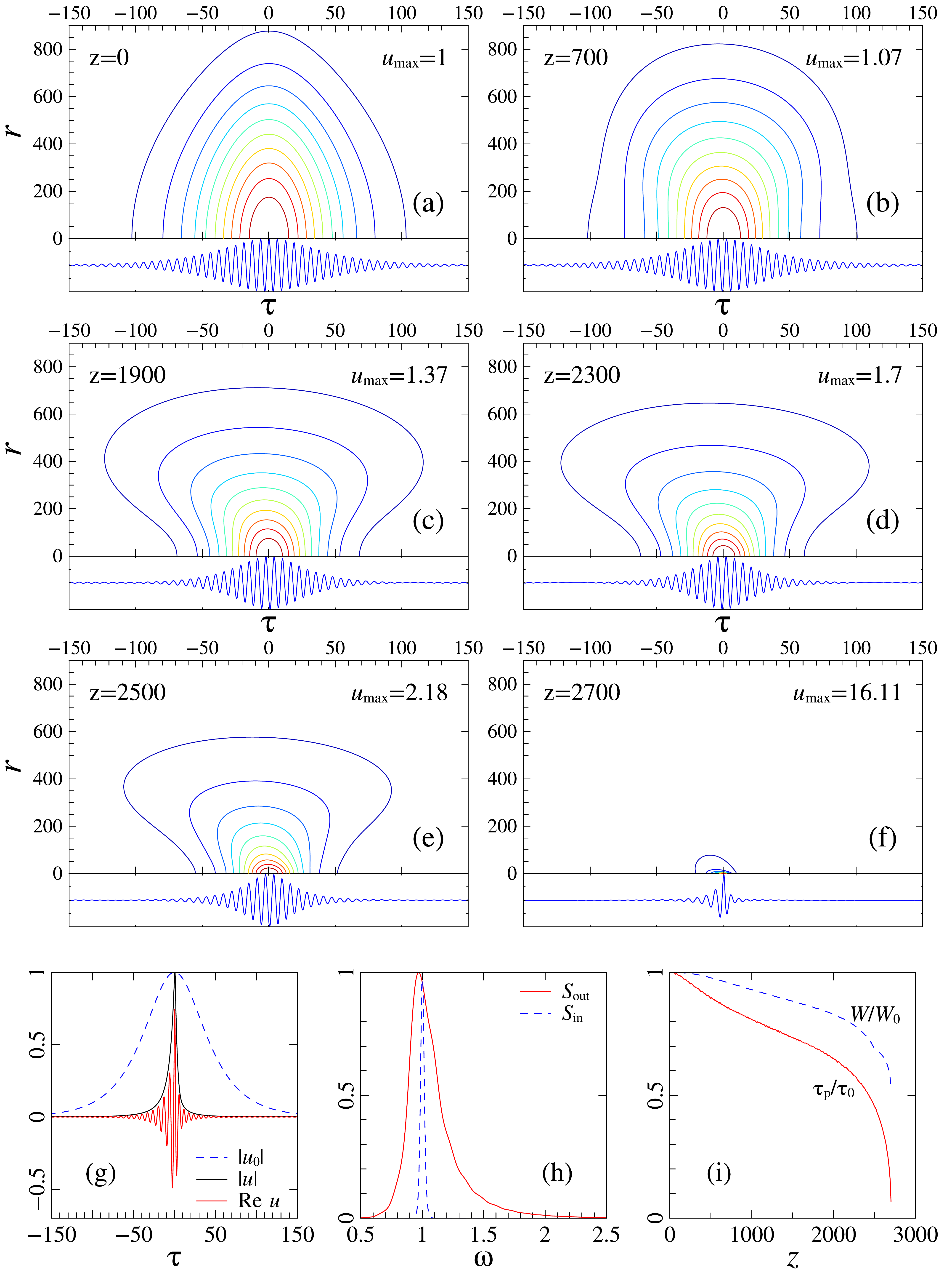}}
	\caption{{\bf (a-f)} Dynamics of circularly polarized field $|u(z,\tau,r)|$ with initial profile Eq.~\eqref{eq:64} for $\mathcal{N}=1$, $\delta=0.03$, $\omega_s=1$, $a=400$.  The \textcolor{blue}{blue line} shows pulse evolution on the beam axis for one of the field components. {\bf (g)} The \textcolor{blue}{blue dashed line} is the distribution of the field envelope of the input laser pulse on the beam axis, and the \textcolor{red}{red line} is the distribution of the field of the compressed pulse on the beam axis. {\bf (h)} \textcolor{blue}{Blue dashed line} is the initial spectrum, \textcolor{red}{red solid line} is the spectrum of compressed pulse. {\bf (i)} Pulse duration and energy of the compressed pulse (red line and blue dashed line, respectively) as functions of the $z$ coordinate. Coordinates $r$, $z$, $\tau$ are dimensionless one according to Eq.~\eqref{eq:5}.}\label{ris:ris4}
\end{figure}

Figure~\ref{ris:ris4} presents the results of numerical simulation with the initial distribution \eqref{eq:64} with $\mathcal{N}=1$, $\delta_0=0.03$, $\omega_s=1$, and $a=400$. In this case, the duration of the wave packet corresponds to ten optical cycles ($\tau_p^{in}=10~ T_0$, where $T_0$ is the field period). It should be emphasized, that the distribution of the laser pulse with the longitudinal scale being much smaller than the transverse one is specified at the input to the nonlinear medium, \ie the dispersion length of the wave packet is much shorter than the diffraction length in the dimensionless equation~\eqref{eq:5}. One can see from Fig.~\ref{ris:ris4}{\bf (a-f)} that self-focusing of the radiation in the transverse direction is accompanied by the adiabatic decrease in the soliton duration. The evolution of the field in the pulse on the beam axis $u_x(r=0)=Re(u)$ is shown as the blue line. It is seen in the figure that the pulse scales decrease significantly. For further adjustment of the numerical simulation results and the above-presented qualitative analysis of the problem, it is convenient to determine the integral pulse width $\rho_\perp$ \eqref{eq:8}. Analysis shows that the average pulse width $\sqrt{\langle\rho_\perp ^2\rangle}$ decreases by $3$~times, from $\sqrt{\langle\rho_\perp ^2\rangle}=400$~to $\sqrt{\langle\rho_\perp ^2\rangle}\simeq133$, and the intensity of the field in the compressed pulse increases by $230$~times.

The dashed line in Fig.~\ref{ris:ris4}{\bf (g)} represents the initial distribution of the pulse envelope~$|u|=\sqrt{u_x^2+u_y^2}$ at the beam axis and the field distribution in the compressed pulse for $u_x=Re(u)$ at the output of the nonlinear medium, $z=2700$. One can see that at the half-intensity level, the laser pulse is compressed by $14$~times, from $\tau_p^{in}=10~ T_0$ to $\tau_p^{out}=0.71~ T_0$, which corresponds to the duration being slightly less than the optical cycle, while the r.m.s. duration of the pulse $\tau_\text{pulse}$, which is calculated on the basis of the second-order momentum, 
\begin{equation}\label{eq:65}
\tau_\text{pulse}=\sqrt{\dfrac{1}{\mathcal{I}_\text{full}} \iint(\tau-\langle\tau\rangle)^2|u|^2rdrd\tau } ,
\end{equation}
amounts to $\tau_\text{pulse}=1.1~ T_0$. 
Here, $\langle\tau\rangle$ is the center of mass of the wave packet. 
\begin{equation}\label{eq:66}
\langle\tau\rangle=\dfrac{1}{\mathcal{I}_\text{full}} \iint |u|^2\tau rdrd\tau .
\end{equation}
It is seen that the pulse duration $\tau_\text{pulse}$ averaged with respect to the transverse distribution of the field intensity, is $1.5$~times larger than the duration of the wave packet pulse at the beam axis. 

Then, we use formula~\eqref{eq:29b} to estimate the duration of the compressed laser pulse. The average wave packet width have decreased by $3$~times. Therefore, the pulse duration determined by using Eq.~\eqref{eq:29b} should decrease to $a_\parallel=1.05~ T_0$, which is a little different from the r.m.s. duration $\tau_\text{pulse}$ of the laser pulse. Therefore, the results of numerical simulation agree with the above-presented qualitative analysis. 

It should be emphasized that in the process of nonlinear dynamics of the laser pulse, the longitudinal scale is always smaller than the transverse one, \ie the distribution of the wave packet is not symmetrized and, therefore, self-focusing of the pulse does not pass over to the regime of spherically symmetric collapse. 

Evidently, such a great decrease in the duration of the laser pulse should be accompanied by a great widening of the wave packet spectrum. The dashed line in Fig.~\ref{ris:ris4}{\bf(h)} shows the spectral intensity of the input pulse at the beam axis, and the solid line is the spectral intensity of the compressed pulse. One can see in this figure that the spectrum of the wave packet at the output from the nonlinear medium is asymmetric and alike the spectrum presented in Fig.~\ref{ris:ris1}{\bf (d)} for the precise soliton solution is found within the framework of the one-dimensional problem~($\Delta_\perp \equiv0$). This asymmetry of the spectrum intensity takes place due to the fact that, as the pulse duration decreases, the term being responsible for steepening of the wave packet profile starts manifesting itself. As it has been already mentioned, in the case of short durations, the soliton solutions have a sufficiently strong frequency modulation (see Eq.\eqref{eq:10a}), which is reflected as significant widening of the short-wave part of the spectrum. 
Note, that the pulse compression in the beam axis is stronger than the average one, since the field intensity is larger in the near-axis region of the pulse.

The solid red line in Fig.~\ref{ris:ris4}{\bf (i)} shows the dependence of the pulse duration, which is determined at the half-maximum of the intensity normalized with respect to the initial value, on the evolution variable~$z$. This plot indicates that the dependence of the wave packet duration has two scales. It follows, in particular, from the system of equations for the wave packet duration \eqref{eq:29b} within the framework of NSE. In the case, where the wave packet width is much smaller than the initial value ($a_\perp \ll a_{\perp 0}$), which corresponds to a significant decrease in the wave packet duration, the behavior of the pulse duration is described by Eq.~\eqref{eq:32b}. Let's consider now the case, where the pulse width is slightly smaller, $a_\perp -a_{\perp0}=\delta a_\perp \ll1$, which corresponds to an insignificant decrease in the wave packet duration. 

The dashed line in Fig.~\ref{ris:ris4}{\bf (i)} shows the dependence of the part of the initial energy in the compressed pulse on the evolution variable $z$. It is seen from the figure presented that the compressed wave packet contains more than 55\% of the initial energy and, since the pulse duration is ten times shorter, the peak power of the compressed pulse increases by 5 times. The remaining energy is located in the halo around the central peak.

\subsection{Multi-soliton dynamics ($\mathcal{N}\geq2$)}\label{sec:sec6_2}
Let's consider now the case of higher $\mathcal{N}$. Our interest in this case is connected with the fact that we can operate with high energies in the laser pulse. This is important in the problem of optimization of the laser pulse self-compression, since the length of the nonlinear medium, at which the laser pulse duration reaches the minimal value, increases in proportion to the increase in the initial transverse width $a$ (Eq.~\eqref{eq:34}). It will be shown in what follows that for $\mathcal{N}>1$, the length of the laser pulse compression will become several times shorter as compared with the initial soliton-like distribution at~$\mathcal{N}=1$. 

Within the one-dimensional problem ($\Delta_\perp \equiv0$), \ie in the absence of spatial effects, the nonlinear dynamics of the laser pulse is determined entirely by the parameter $\mathcal{N}$.  It has been already noted that the initial distribution of the pulse coincides exactly with the soliton solution, Eq.~\eqref{eq:9} in the case of $\mathcal{N}=1$. At $\mathcal{N}\geq2$, the initial wave packet at the asymptotic stage will split into a sequence of solitons with the parameters $\delta_n=(2n-1)\delta_0$, where $n=1,\ldots, [\mathcal{N}]$ is a sequence of integer numbers~\cite{soliton}. Here $[\mathcal{N}]$ is the integer part of $\mathcal{N}$. An important feature of the wave solitons under consideration (Eqs.~\eqref{eq:9} and \eqref{eq:10}) is the semi-bounded spectrum of their permissible solutions, \ie the presence of the bounding parameter~$\delta_\text{cr}$ corresponding to the limiting soliton with the minimum possible pulse duration and, correspondingly, the maximum permissible amplitude. Therefore, the number of solitons, into which the initial pulse splits actually, is an integer number $[\mathcal{N}]$, obeying effectively the following inequality ($\delta^{([N])}<\delta_\text{cr}$):
\begin{equation}\label{eq:69}
\left(2[\mathcal{N}]-1\right)\delta_0<\delta_\text{cr}=\sqrt{\dfrac18} .
\end{equation}
The pulse dynamics will be more complicated, but in the long run, the solitons with $\delta^{([N])}<\delta_\text{cr}$ will be formed. 

\begin{figure}[htpb]
	\center{\includegraphics[width=\linewidth]{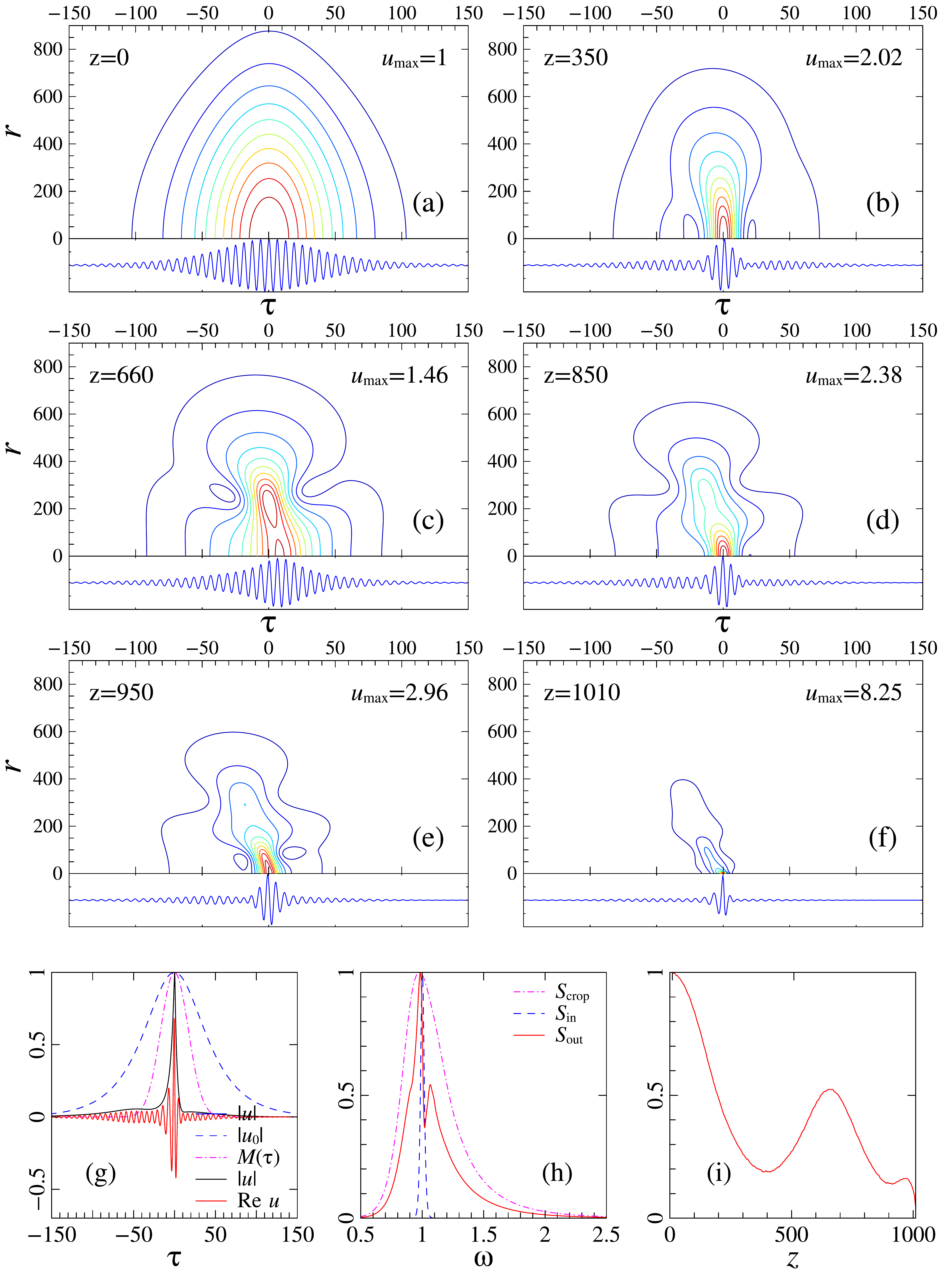}}
	\caption{{\bf (a-f)} 
		Dynamics of the circularly polarized field $|u(z,\tau,r)|$ for $\mathcal{N}=2.05$, $\delta_0=0.03$, and $\omega_s=1$ with the Gaussian distribution in the transverse direction with width $a=400$.
		The \textcolor{blue}{blue line} in inplots shows the pulse evolution at the beam axis ($r=0$) for one of the field components, $u_x=Re(u)$. Here, the field is normalized with respect to the maximum value.
		The {\bf (g)} \textcolor{blue}{dashed blue line} is the distribution of the field envelope of the input laser pulse at the beam axis~$|u(\tau,r=0)|$, the 
		\textcolor{red}{red line} is the distribution of the compressed-pulse field at the beam axis$u_x=Re(u)$, the 
		\textcolor{magenta}{dash-and-dotted magenta line} is the distribution of the time mask, the 
		\textcolor{magenta}{dashed magenta line} is the distribution of the spectrum intensity after application the time mask over the compressed pulse, the 
		{\bf (h)} \textcolor{blue}{dashed blue line} is the initial spectrum, the 
		\textcolor{red}{red line} is the spectrum of the compressed pulse, and the 
		{\bf (i) } \textcolor{red}{red line} is the dependence of the wave packet duration normalized with respect to the initial duration for $z$.}\label{ris:ris6}
\end{figure}

We turn now to the initial problem and analyze the dynamics of the laser pulse (Eq.~\eqref{eq:64}) within the framework of Eq.~\eqref{eq:5} allowing for the spatial effects. Figure~\ref{ris:ris6}{\bf (a-f)} presents the spatio-temporal evolution of the laser pulse at $\mathcal{N}=2.05$ and $a=400$, when the distribution was specified in the longitudinal direction at $\delta=0.03$ and $\omega_s=1$. It is seen from this figure that at the initial stage ($z\simeq 350$), the laser pulse is strongly compressed according qualitative law \eqref{eq:29b}. At this the pulse width decreases not so strongly.

Then, just as within the framework of the one-dimensional problem, the wave packet starts blurring in the longitudinal direction ($z\sim 660$), which leads to a decrease in the pulse self-focusing rate, since the amplitude of the field has decreased. As seen from the figure, at $z= 660$, a horseshoe-shaped structure starts forming. Then, as follows from the figure, at $z\sim 850$, the pulse starts splitting in the longitudinal direction into two solitons. The soliton with a short duration and, hence, a high amplitude, is located at the rear front of the pulse ($\tau \in [-25,25]$), and the soliton with a greater duration and, hence, a lower amplitude is located at the leading front of the time distribution ($\tau \in [-150,0]$). Note that in the near-axis part of the pulse the field becomes again stronger, than in the edge region~$r\sim a/2$. Further, due to the process of pulse self-focusing, the duration of these solitons will decrease adiabatically. Evidently, the short soliton will be compressed faster, since its energy is greater, $\mathcal{I}_2\simeq12\pi\delta_0 a_0^2/a_\text{cur}^2$, as compared with the longer one $\mathcal{I}_1\simeq4\pi\delta_0 a_0^2/a_\text{cur}^2$, where $a_\text{cur}$ is the current beam width, and $a_0$ is the initial beam width. As a result of the laser beam self-focusing, a soliton with a shorter duration will be segregated which is shown in Fig.~\ref{ris:ris6}{\bf (a-f)} for $z\sim1010$. 

The blue dashed line in Fig.~\ref{ris:ris6}{\bf (g)} shows the initial distribution of the envelope of the pulse $|u|=\sqrt{u_x^2+u_y^2}$ at the beam axis and the red line shows the distribution of the field in the compressed pulse for $u_x=Re(u)$ at the output of the nonlinear medium, $z=1010$. It is seen from the figure that the laser pulse at the half-maximum of the field intensity is compressed by $19$ times, from $\tau_p^{in}=10~ T_0$ to $\tau_p^{out}=0.52~ T_0$, which corresponds to a value being slightly less than the optical cycle. As is seen in the figure, a long soliton, which is almost unnoticeable and is represented by a small pedestal at the level $0.02$, is located at the leading edge of the time distribution of the laser pulse $\tau \in [-125,-25]$. Therefore, this pedestal will be totally unobservable  in the intensity distribution against the background of the main signal. 

Such a significant shortening of the output pulse duration, as it has been already mentioned, should be accompanied with a significant widening of the spectrum of the compressed pulse. The blue dashed line in Fig.~\ref{ris:ris6}{\bf (h)} represents spectral intensity of the input pulse at the beam axis, and the solid red line shows the spectral intensity of the compressed pulse. Note that the ruggedness of spectral intensity of the laser pulse at the output of the nonlinear medium $z\simeq 1010$ appears due to the interference of two temporally separated wave structures. 

To develop spectral intensity of just one soliton with a short duration, we applied the time-interval mask $\mathcal{M}(\tau)$
\begin{equation}\label{eq:70}
\mathcal{M}(\tau)=\exp\left[-2\log2\left(\dfrac{\tau}{28}\right)^2 \right] 
\end{equation}
to the compressed pulse, in order to remove the soliton with a longer duration, which is located on the leading edge. The distribution of the time-interval value $\mathcal{M}(\tau)$, Eq.~\eqref{eq:70} is shown in Fig.~\ref{ris:ris6}{\bf (g)} as a dash-and-dotted magenta line, whereas the dash-and-dotted magenta line in Fig.~\ref{ris:ris6}{\bf (h)} rerpresents spectral intensity of the output laser pulse at the beam axis after application of the time-interval value $\mathcal{M}(\tau)$. As seen in Fig.~\ref{ris:ris6}{\bf (i)}, the resulting spectrum has become smooth and more asymmetric, and looks like the spectra shown in Figs.~\ref{ris:ris1}{\bf (d)} and \ref{ris:ris4}{\bf (h)}. In this case, the spectral intensity of the short soliton is wider than the spectral intensity of the compressed laser pulse for the case, where only one soliton contains in the initial wave packet ($\mathcal{N}=1$). 

The solid red line in Fig.~\ref{ris:ris6}{\bf (i)} shows the dependence of the pulse duration, which is determined at half-maximum of the field intensity normalized with respect to the initial value $\tau_0$, on the evolution variable $z$. This figure demonstrates two stages in the evolution of a laser pulse, which we discussed earlier.
Specifically, as follows from Fig.~\ref{ris:ris6}{\bf (i)}, the duration of the wave packet reaches the intermediate minimum $\tau_p\simeq0.18 \tau_0$ at $z\simeq380$.
Presumably, one can restrict consideration to this medium length for compression of the initial laser pulse. However, in this case, self-compression of the wave packet is rather sensitive to the length of the nonlinear medium, since, as seen in Fig.~\ref{ris:ris6}{\bf (i)}, at $z\sim400$ the pulse duration starts increasing again, just as within the framework of the one-dimensional problem, and reaches $\tau_p=0.52~ \tau_0$ at $z\simeq650$. It has been already mentioned that the pulse splits into two solitons, which start compressing adiabatically. 
So, from the practical point of view, it is preferable to compress a pulse at $z\gtrsim 800$ due to following reasons. First, the amplitude of the tailing soliton exceeds significantly the amplitude of the leading soliton. Second, the decrease in the duration of the wave packet occurs monotonically, as the pulse propagates in the medium.

Note that at $z\simeq1010$, the duration of the laser pulse is $3.9$~times shorter, than at $z\simeq380$. As it follows from the comparison of Figs.~\ref{ris:ris4}{\bf (i)} and \ref{ris:ris6}{\bf (i)}, the initial laser pulse at $\mathcal{N}=2.05$ compresses along the path $z\simeq1010$, which is significantly shorter, than the path $z\simeq2700$ for the case $\mathcal{N}=1$. It should be noted that such a significant decrease in the length of the wave packet compression happens due to the pulse splitting into two structures, the duration of a shorter pulse decreases by three times compared with the initial duration, which in the long run will lead to a decrease in the compression length. 

\begin{figure}[htpb]
	\center{\includegraphics[width=\linewidth]{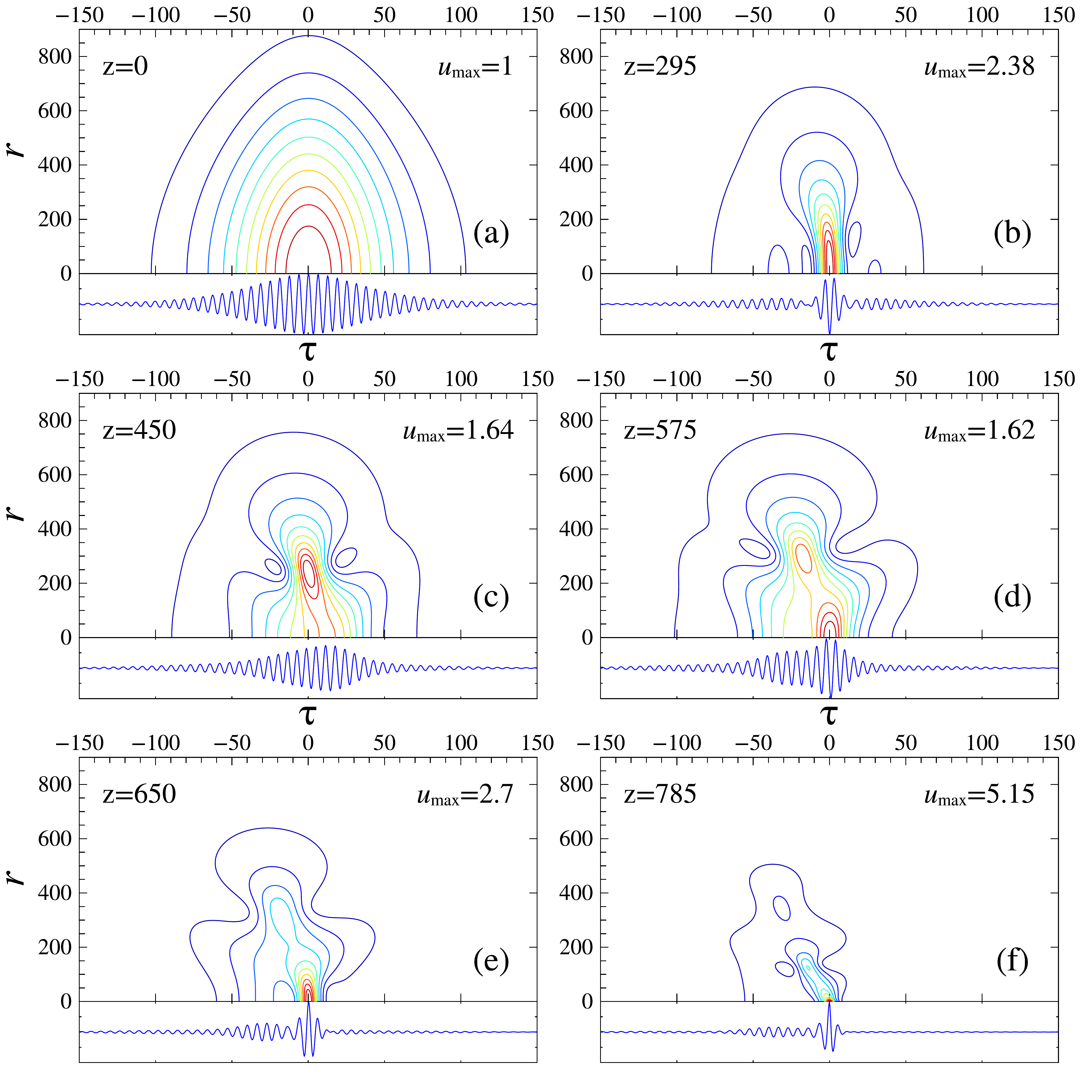}}
	\caption{Dynamics of the circularly polarized field $|u(z,\tau,r)|$ for $\mathcal{N}=2.5$, $\delta_0=0.03$, and $\omega_s=1$ with the Gaussian distribution in the transverse direction with width $a=400$.
		The \textcolor{blue}{blue line} in inplots shows the pulse evolution at the beam axis ($r=0$) for one of the field components, $u_x=Re(u)$. Here, the field is normalized with respect to the maximum value. Coordinates $r$, $z$, $\tau$ are dimensionless according to Eq.~\eqref{eq:5}.}\label{ris:ris7}
\end{figure}

\begin{figure}[htpb]
	\center{\includegraphics[width=\linewidth]{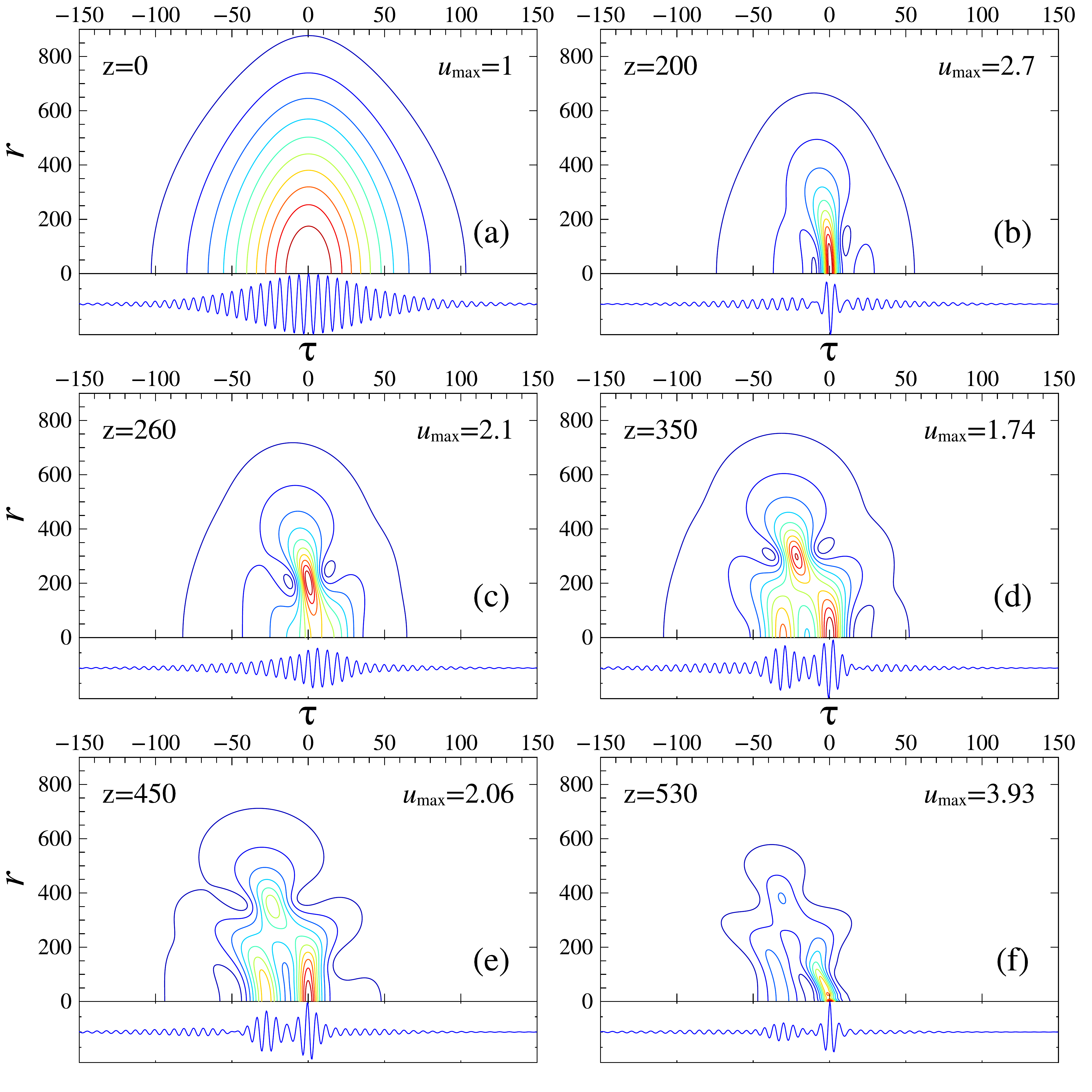}}
	\caption{The same as figure \ref{ris:ris7} but with $\mathcal{N}=3.02$.}\label{ris:ris7b}
\end{figure}

In the process of the further increase of the parameter $\mathcal{N}$, self-compression of the laser pulse in the process of self-focusing of the transverse distribution is preserved (Fig.~\ref{ris:ris7},~\ref{ris:ris7b}). The figure presents the evolution of the laser pulse for two different values of the parameter $\mathcal{N}$: {\bf (a)} -- $\mathcal{N}=2.5$, {\bf (b)} -- $\mathcal{N}=3.02$. It follows from the figure that at $z\sim650$ {\bf (a)}, the laser pulse splits into two wave structures, and the soliton with a lower amplitude becomes more pronounced in the background of a shorter-duration soliton, unlike the case shown in Fig.~\ref{ris:ris6}{\bf (a-f)}. At $\mathcal{N}=3.02$, the laser pulse splits into three structures at $z\sim450$ (see Fig.~\ref{ris:ris7b}). It is evident that for self-compression of the input laser pulse, it is desirable to restrict consideration to the length of the nonlinear medium being $z\sim295$ for $\mathcal{N}=2.5$ and $z\sim200$ for $\mathcal{N}=3.02$, since further the time structure of the laser pulse becomes rather complicated. 

Thus, in the case of self-compression of laser pulse under the conditions of self-focusing of spatial distribution, when the linear dispersion length is much shorter than the diffraction length, the key role in the pulse dynamics is played by the solitons found within the framework of the one-dimensional problem~\cite{soliton}. In the case of $\mathcal{N}>1$, the prevalent role is played by the modulation instability of the wave field, rather than the filamentation one. It leads to splitting of the laser pulse into a set of soliton-like structures~\cite{soliton}, which further, due to the process of self-focusing, will compress individually and monotonically in the longitudinal direction. As shown by the numerical analysis, in order to obtain extremely short laser pulses with high time contrast, it is preferable to specify wave field distribution~\eqref{eq:64} with $N\lesssim2.5$ to the input of the nonlinear medium.

\section{Conclusions}

In this paper, we justify theoretically the promising method of self-compression of multi-millijoule laser pulses up to one optical cycle. Self-focusing of a wave field in a medium with the Kerr-type inertia-free nonlinearity and anomalous dispersion of group velocity leads to an adiabatic decrease of the wave packet duration to a duration comparable with the optical cycle for wide wave packets with the soliton-like field distribution along the longitudinal coordinate. Analysis shows that the soliton duration decreases in proportion to the square of characteristic width of the wave beam~\eqref{eq:29b}. It should be noted that the strongly oblate ellipsoidal distribution of the wave beam is preserved in the process of evolution, and no symmetrization occurs. Self-compression of the laser pulse proceeds under the conditions of a noticeable excess over the threshold value of the self-focusing power.

Thorough numerical studies of the evolution of a 3D axisymmetric wave packet have been performed. The results obtained on the basis of qualitative analysis in the aberration-free approximation have been confirmed by the numerical simulation. In the case of the quasi-soliton field distribution in the longitudinal direction, the pulse self-focusing is accompanied by a monotonic decrease of its duration down to the one optical cycle, corresponding to the duration of the limiting soliton \eqref{eq:9}.

In the case of the initial high amplitude of the laser pulse ($\mathcal{N} \gtrsim 2$), the initial longitudinal distribution of the wave packet splits into a sequence of solitons, which further self-compress monotonically and diverge in the longitudinal direction. Numerical simulations show that it is preferable to use wave packets containing not more than two quasi-soliton structures. 

We should denote that self-action of intense laser pulses having durations shorter than ten optical cycles is stable relative to the transverse filamentation instability \cite{Skobelev16}. This takes place due to allowance for the nonlinear medium dispersion (dependence of the group velocity on the amplitude), under which the type of self-focusing instability changes from the absolute to the convective one. As a result, in the case of sufficiently short pulses, the noise amplitudes shift relative to the pulse and have no time to reach to arbitrary noticeable values. 

In recent paper~\cite{Kartashov}, the possibility of laser pulse self-compression at a wavelength of $3{.}9$~$\mu$m from $94$~fs to $30$~fs was demonstrated for an input radiation power exceeding the critical self-focusing power by four orders of magnitude. The duration of the  wave packet decreased by three times after it had passed through a YAG plate having a thickness of $2$~mm. Simple evaluations show that the plate thickness was chosen to be less than the length, at which the filamentation instability develops, which leads to decomposition of the transverse field distribution to separate filaments. However, our study shows that much stronger compression down to 7~fs pulse duration (about optical cycle) can be achieved under the similar conditions, but with 7\ldots8 times larger medium length.

Let us place some estimates for the realistic experimental realization of proposed laser pulse compression. The typical initial parameters of an appropriate laser pulse are: 	the wavelength of 4$\mu$m, the initial duration $\tau_\text{in} = 100$~fs, the energy $W_\text{in} = 15$~mJ, the initial beam radius of $0.15$~cm. This corresponds to the power $P_\text{in} = 150$~GW. At this typical critical self-focusing power is $P_\text{cr} \simeq 20$~MW for this frequency range (for example, YAG plate has such critical power). Most media in this frequency range have anomalous group velocity dispersion, which is a prerequisite for self-compression of the wave packet in the process of the radiation self-focusing. As a result, the laser pulse duration will decrease to $\tau_\text{out} \simeq 7$~fs, and pulse energy will become about $W_\text{out} \simeq 7$~mJ ($P_\text{out} = 1$~TW) after passing YAG plate with thickness of 1.5~cm. This case corresponds to the pulse duration smaller than initial field period.

This work was supported by the Russian Science Foundation (Project No. 16-12-10472) and RFBR (Project No. 15-32-20641).

\end{document}